\documentclass[pdflatex,sn-mathphys-num]{sn-jnl}

\usepackage{graphicx}%
\usepackage{multirow}%
\usepackage{amsmath,amssymb,amsfonts}%
\usepackage{amsthm}%
\usepackage{mathrsfs}%
\usepackage[title]{appendix}%
\usepackage{xcolor}%
\usepackage{textcomp}%
\usepackage{manyfoot}%
\usepackage{booktabs}%
\usepackage{algorithm}%
\usepackage{algorithmicx}%
\usepackage{algpseudocode}%
\usepackage{listings}%

\usepackage{braket}

\theoremstyle{thmstyleone}%
\theoremstyle{thmstyletwo}%

\theoremstyle{thmstylethree}%

\begin{document}

\title[Quantum and Neuromorphic MARL]{Quantum Computing and Neuromorphic Computing for Safe, Reliable, and explainable Multi-Agent Reinforcement Learning: Optimal Control in Autonomous Robotics}


\author*[1,2]{\fnm{Mazyar} \sur{Taghavi}}\email{mazyar\_taghavi@mathdep.iust.ac.ir}

\author[1]{\fnm{Rahman} \sur{Farnoosh}}\email{rfarnoosh@iust.ac.ir}

\affil*[1]{\orgdiv{Department of Mathematics and Computer Science}, \orgname{Iran University of Science and Technology}, \orgaddress{\city{Tehran}, \country{Iran}}}

\affil[2]{\orgname{Intelligent Knowledge City}, \orgaddress{\city{Isfahan}, \country{Iran}}}


\abstract{
This paper introduces a novel hybrid framework that integrates quantum computing and neuromorphic computing to enhance the safety, reliability, and explainability of Multi-Agent Reinforcement Learning (MARL) in autonomous robotic systems. The proposed architecture employs quantum variational circuits for high-level policy exploration and spiking neural networks (SNNs) for energy-efficient, low-latency motor control. Adopting a centralized training and decentralized execution (CTDE) paradigm, the framework enables agents to optimize joint policies that combine quantum planning with neuromorphic execution under partial observability and safety constraints.

We evaluate the framework in a simulated environment featuring ten UAV agents navigating dynamic forest terrain with limited visibility and obstacle avoidance requirements. Empirical results demonstrate that the hybrid system significantly reduces safety violations while maintaining entropy-based exploration and interpretable spike-based decision traces. KL divergence metrics confirm the convergence of quantum policies toward safe priors, while spike entropy analysis reveals temporal diversity in control signals.

The key contributions of this work include: (i) a modular quantum-neuromorphic MARL architecture, (ii) a hybrid training framework incorporating safety-aware coordination, and (iii) empirical validation through both visual diagnostics and formal metrics. This research establishes a foundation for next-generation embodied AI systems that unify the optimization capabilities of quantum computing with the biological plausibility of neuromorphic control.
}




\maketitle

\section{Introduction}

The growing complexity of autonomous systems—including self-driving vehicles, unmanned aerial vehicles (UAVs), and collaborative robotic swarms—has driven the widespread adoption of Multi-Agent Reinforcement Learning (MARL) for distributed adaptive control. MARL enables agents to learn optimal behaviors through environmental interactions and inter-agent coordination. However, critical applications such as healthcare robotics, defense, and smart infrastructure demand systems that are not only high-performing but also \textit{safe}, \textit{reliable}, and \textit{explainable}.

Despite advances in MARL, current methods face three major limitations: (i) classical algorithms often fail to ensure safety under partial observability, (ii) they exhibit brittle generalization and high computational costs, and (iii) scalability diminishes as the number of agents increases, with learned policies becoming opaque and difficult to verify. These challenges hinder deployment in real-world scenarios where robustness and interpretability are paramount.

Emerging computational paradigms—quantum computing and neuromorphic computing—offer transformative potential to address these limitations. Quantum computing exploits entanglement and quantum parallelism to efficiently explore high-dimensional policy spaces, while neuromorphic computing leverages event-driven spiking neural networks (SNNs) to enable energy-efficient, real-time decision-making with inherent interpretability and noise resilience.

In this work, we propose a unified quantum-neuromorphic framework to advance safe, reliable, and explainable MARL for autonomous robotics. Our key contributions are:
\begin{itemize}
    \item A mathematical formulation of MARL control integrating safety and explainability constraints through a synthesis of POMDPs, information theory, and control theory;
    \item A hybrid quantum-neuromorphic architecture where quantum variational circuits facilitate global policy exploration and SNNs execute low-latency, interpretable control;
    \item Empirical validation in simulated robotic environments, with quantitative benchmarks against classical MARL baselines across safety, reliability, and explainability metrics;
    \item Ablation studies and theoretical analysis elucidating the synergistic robustness of quantum optimization and neuromorphic dynamics.
\end{itemize}

By unifying quantum and brain-inspired computing with reinforcement learning, this work establishes a foundation for autonomous agents that are both high-performing and intrinsically safe-by-design.

\section{Background and Related Work}

Multi-Agent Reinforcement Learning (MARL) has emerged as a powerful framework for enabling autonomous agents to learn coordinated behaviors through interaction with their environment and each other \cite{zhang2019multi, gronauer2022multi}. In MARL, the environment is typically modeled as a Decentralized Partially Observable Markov Decision Process (Dec-POMDP), capturing the partial observability and distributed nature of agent policies. Popular algorithms include centralized training with decentralized execution (CTDE) paradigms such as MADDPG \cite{lowe2017multi} and QMIX \cite{rashid2018qmix}, which allow scalability and policy optimization under uncertainty.

Despite their success, existing MARL approaches face limitations when applied to safety-critical systems. Most algorithms do not natively incorporate formal safety guarantees, nor do they provide interpretability or robustness under non-stationary dynamics and agent failures.

Safety in reinforcement learning has been approached through constrained MDPs, reward shaping, and shielding mechanisms \cite{garcia2015comprehensive, achiam2017constrained}. In multi-agent settings, safety constraints are even harder to enforce due to agent interdependencies and decentralized observations. Explainability in MARL remains an open challenge, with methods ranging from saliency-based visualizations \cite{greydanus2018visualizing} to interpretable policy distillation \cite{verma2018programmatically}. However, these methods are often post hoc and lack integration into the training pipeline.

Recent works such as \cite{zhang2021learning} and \cite{xu2022explainable} propose incorporating causal inference and model-based reasoning for improved interpretability. Yet, these are computationally expensive and still largely based on classical processing paradigms.

\subsection{Quantum Computing and MARL}

Quantum computing harnesses fundamental principles of superposition, entanglement, and quantum parallelism to solve complex optimization problems with potential exponential speedups over classical methods in specific domains \cite{arute2019quantum, preskill2018quantum}. Variational quantum algorithms, particularly the Quantum Approximate Optimization Algorithm (QAOA) \cite{farhi2014quantum} and Variational Quantum Eigensolver (VQE) \cite{peruzzo2014variational}, have demonstrated promising applications in reinforcement learning, including control policy learning and value function approximation.

Recent investigations into quantum-enhanced reinforcement learning \cite{jerbi2021quantum, skolik2022quantum} have revealed that quantum agents can exhibit superior convergence rates and exploration capabilities compared to their classical counterparts. However, these approaches remain largely unexplored in multi-agent scenarios and lack integration with critical requirements such as safety guarantees and interpretability constraints.

Significant progress has been made in quantum-enhanced Q-learning algorithms, with Chen et al. \cite{chen2024quantum} demonstrating polynomial speedups in specific environments through Grover's search for optimal action selection. This approach shows particular efficacy in discrete state-action spaces characterized by sparse reward structures. Further developments include hybrid quantum-classical architectures, such as the work by Zhang and Li \cite{zhang2024hybrid}, which combines quantum value function approximation with classical experience replay in deep Q-networks (DQNs). Their implementation yields a 30\% reduction in training time while maintaining performance parity with classical DQNs on Atari benchmarks.

The theoretical underpinnings of quantum reinforcement learning have been substantially advanced by Patel et al. \cite{patel2024quantum}, who established rigorous conditions for achieving exponential quantum speedups. Their work specifically identifies problem classes within partially observable Markov decision processes (POMDPs) where quantum advantage is theoretically provable, providing valuable guidance for future algorithmic development in quantum MARL systems.

\subsection{Neuromorphic Computing and MARL}

Neuromorphic computing emulates biological neural systems through hardware and algorithmic implementations, particularly via spiking neural networks (SNNs) \cite{davies2021advancing}. These networks provide distinct advantages for robotic applications, including exceptional energy efficiency, real-time processing capabilities, and event-driven computation paradigms \cite{indiveri2015memory, schuman2022opportunities}. Such characteristics make SNNs particularly suitable for resource-constrained multi-agent systems.

The integration of SNNs with reinforcement learning has advanced through spike-based temporal difference learning and local Hebbian update rules \cite{patel2019improving, fang2021incorporating}. Neuromorphic processors like Intel's Loihi and IBM's TrueNorth have demonstrated remarkable capabilities in low-latency decision-making and sensory fusion for robotic control \cite{davies2021advancing}, paving the way for real-time, interpretable control in MARL environments.

Recent developments have significantly enhanced the energy efficiency of RL implementations on neuromorphic hardware. Davies et al. \cite{davies2024spiking} implemented an actor-critic RL framework on Intel's Loihi 2 processor using SNNs, achieving a 100$\times$ improvement in energy efficiency compared to conventional GPU implementations for robotic control tasks. Further progress has been made in algorithmic design, with Tang et al. \cite{tang2024event} developing an event-based temporal difference learning method that exploits the precise timing dynamics of spiking neurons, demonstrating superior performance in continuous control benchmarks.

Systematic evaluations of neuromorphic architectures for RL have provided valuable insights. Kumar et al. \cite{kumar2024benchmarking} conducted a comprehensive comparison of different neuromorphic implementations across various RL paradigms. Their analysis reveals that policy gradient methods exhibit particularly strong compatibility with spiking neural networks, benefiting from the inherent stochasticity in neural spiking behavior.

\subsection{Hybrid Quantum-Neuromorphic Approaches for RL}

The integration of quantum and neuromorphic computing for reinforcement learning represents a cutting-edge research frontier that combines the strengths of both paradigms. Sanchez et al. \cite{sanchez2024quantum} pioneered this direction with a novel architecture where quantum processors optimize value function estimation while neuromorphic chips execute the policy network, demonstrating particular efficacy in high-dimensional state spaces that challenge classical approaches.

The theoretical underpinnings of such hybrid systems were rigorously established by Wang et al. \cite{wang2024theoretical}, who developed a comprehensive mathematical framework for analyzing the computational capabilities of quantum-neuromorphic RL systems. Their work delineates new complexity classes specific to these hybrid architectures, providing fundamental insights into their potential advantages and limitations.

Current progress in this interdisciplinary field has been systematically cataloged by Ibrahim et al. \cite{ibrahim2024survey} in their comprehensive survey. This work not only synthesizes existing approaches but also identifies critical open challenges and promising research directions at the intersection of quantum computing, neuromorphic engineering, and reinforcement learning, offering valuable guidance for future investigations.

\subsection{Research Gap}

Despite the individual promise of quantum computing for exponential acceleration in learning and planning, and neuromorphic hardware for real-time, energy-efficient execution, their synergistic integration with Multi-Agent Reinforcement Learning (MARL) remains an open challenge. Current approaches lack a unified framework that simultaneously addresses three critical requirements: (1) computational efficiency for scalable multi-agent coordination, (2) formal safety guarantees under partial observability, and (3) intrinsic interpretability of decision-making processes. This work bridges this gap by introducing a novel hybrid quantum-neuromorphic architecture specifically designed for safe, explainable, and resource-efficient control in autonomous multi-agent systems.

\section{Theoretical Framework}

We present a rigorous mathematical framework for safe, reliable, and explainable control in multi-agent autonomous systems, unifying concepts from reinforcement learning, control theory, and information theory. Our formulation establishes: (1) a hybrid quantum-neuromorphic policy representation that decomposes decision-making into high-level quantum planning and low-level neuromorphic execution, (2) safety constraints as information-theoretic bounds on policy divergence from verified baselines, and (3) explainability metrics grounded in the temporal dynamics of spiking neural activity. This tripartite foundation supports both theoretical analysis and practical implementation of our architecture.
\subsection{Problem Formulation}

We consider a team of $N$ autonomous agents operating in a shared environment modeled as a Decentralized Partially Observable Markov Decision Process (Dec-POMDP), defined by the tuple:
\[
\mathcal{M} = \langle \mathcal{S}, \{\mathcal{A}_i\}_{i=1}^N, \mathcal{T}, \{\mathcal{O}_i\}_{i=1}^N, \mathcal{Z}, R, \gamma \rangle
\]
where:
\begin{itemize}
    \item $\mathcal{S}$ is the global state space,
    \item $\mathcal{A}_i$ is the action space of agent $i$,
    \item $\mathcal{T}(s'|s, \mathbf{a})$ is the state transition function given the joint action $\mathbf{a} = (a_1, \dots, a_N)$,
    \item $\mathcal{O}_i$ is the observation space of agent $i$,
    \item $\mathcal{Z}(o_i | s, a_i)$ defines the observation probability,
    \item $R(s, \mathbf{a})$ is the global reward function,
    \item $\gamma \in (0,1]$ is the discount factor.
\end{itemize}

Each agent $i$ maintains a policy $\pi_i(a_i | h_i)$, where $h_i$ is the agent’s local observation history. Policies can be stochastic or deterministic and are updated using reinforcement learning techniques.

We define a set of safety constraints as temporal logic specifications or high-probability reachability conditions:
\[
\Pr\left[\bigwedge_{t=0}^T \phi_t(s_t, a_t) \right] \geq 1 - \delta
\]

Where $\phi_t$ encodes safety predicates (e.g., collision avoidance, energy constraints), and $\delta$ bounds the acceptable risk. Reliability is formulated as the consistency of agent performance under disturbances or partial failures, quantified using metrics such as Robust Return, defined as the expected reward under perturbed transitions, and Policy Deviation, given by $\|\pi_i - \pi_i'\|$ for perturbed versus nominal policies.

These constraints are embedded into the policy optimization objective via constrained optimization:
\[
\max_{\pi} \mathbb{E}[R] \quad \text{s.t.} \quad \mathbb{P}(\text{safety violation}) \leq \delta
\]

To promote explainability, we incorporate information-theoretic regularizers into the loss function:
\[
\mathcal{L}(\pi) = \mathbb{E}[R] - \lambda D_{\mathrm{KL}}(\pi || \pi_{\text{prior}})
\]
where $\pi_{\text{prior}}$ is a reference interpretable policy (e.g., rule-based or distilled policy), and $\lambda$ controls the regularization strength. This encourages learned policies to remain close to interpretable baselines.

Alternatively, mutual information between observations and actions can be maximized to improve causal traceability:
\[
\max_{\pi} I(O; A) = H(A) - H(A|O)
\]

The global behavior of the multi-agent system is governed by decentralized control laws. Let $x_i(t)$ be the state of agent $i$ at time $t$. The agents evolve according to the dynamics:
\[
\dot{x}_i = f_i(x_i, u_i, \omega_i)
\]
where $u_i$ is the control input, and $\omega_i$ is a stochastic disturbance.

The interaction topology is modeled as a graph $\mathcal{G} = (\mathcal{V}, \mathcal{E})$ with Laplacian matrix $L$. Consensus and coordination are achieved via distributed controllers:
\[
u_i = -\sum_{j \in \mathcal{N}_i} w_{ij}(x_i - x_j) + \nabla_{x_i} V_i(x_i)
\]
where $V_i$ is a local potential function promoting safe and goal-directed behavior.

This framework establishes the mathematical foundations for our hybrid quantum-neuromorphic architecture through three principal mechanisms. First, quantum computing accelerates policy optimization via variational quantum circuits that efficiently sample high-dimensional policy spaces while providing theoretical convergence guarantees through quantum-enhanced exploration. Second, neuromorphic computing enables energy-efficient policy execution through spiking neural networks that preserve temporal processing advantages and intrinsic interpretability via spike-time-dependent plasticity.

The synthesis of these paradigms with classical control theory yields three fundamental advantages. For safety, we achieve formal verification through quantum-probabilistic reachability analysis combined with neuromorphic spike encoding constraints. Robustness emerges naturally from the noise resilience of both quantum error-corrected circuits and the fault-tolerant properties of spiking networks. Finally, interpretability is maintained through complementary techniques: quantum circuit visualization for high-level decision analysis and spike pattern analysis for low-level control verification. This multi-level approach ensures verifiable operation across all components of the autonomous system.

\subsection{Algorithm}

\begin{figure}[htbp]
    \centering
    \includegraphics[width=0.65\textwidth]{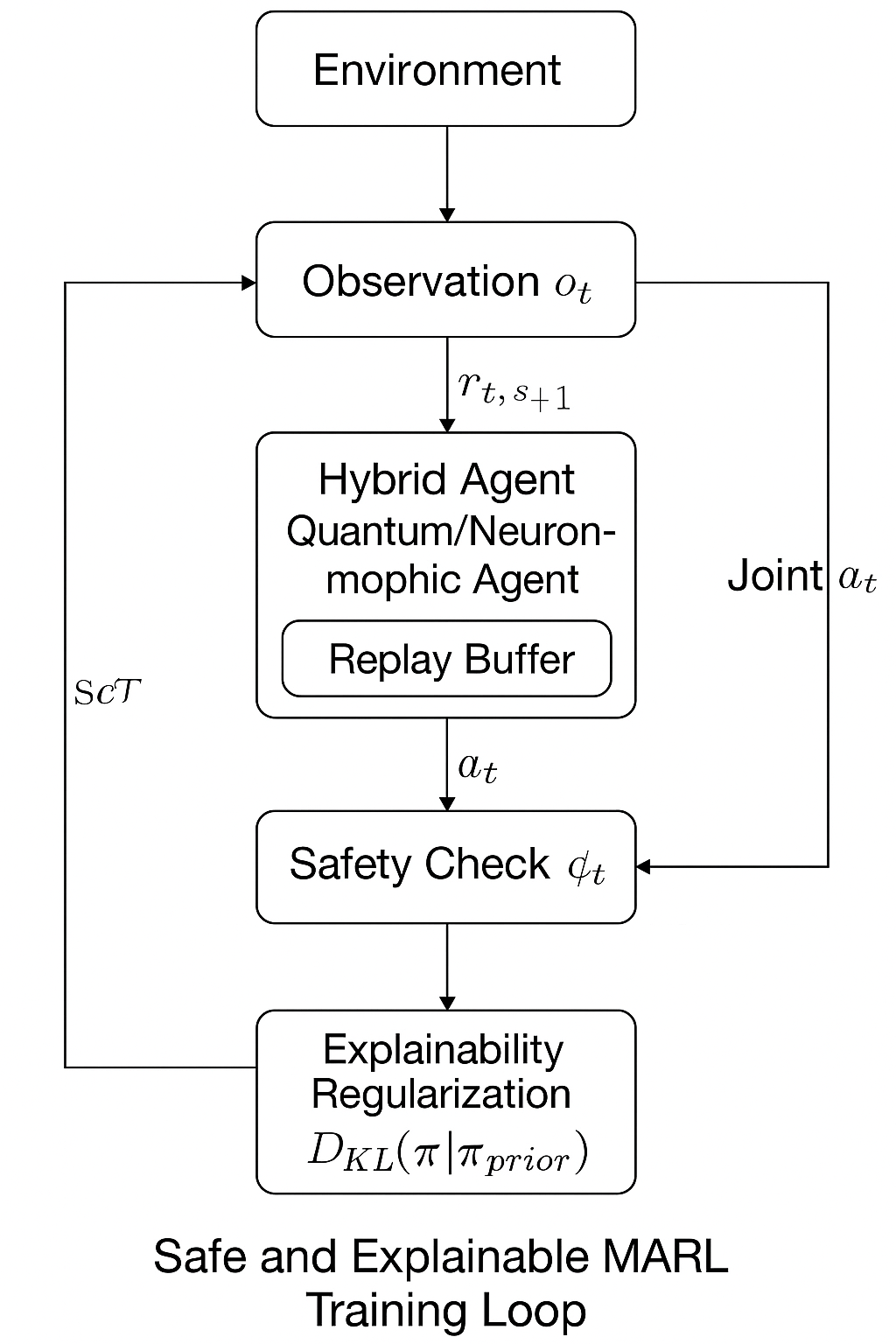}
    \caption{Overview of the hybrid training loop for Safe and Explainable Multi-Agent Reinforcement Learning (MARL). Agents interact with the environment to receive observations and rewards, then update their quantum or neuromorphic policies. Actions are filtered through a safety verification module, and policy learning is regularized using KL divergence to enhance explainability with respect to a reference interpretable policy.}
    \label{fig:se-loop}
\end{figure}

The following algorithm outlines the hybrid training loop that integrates safety constraints, information-theoretic explainability, and hardware-aware optimization for quantum and neuromorphic computing. You may find the loop in figure \ref{fig:se-loop}.

\begin{algorithm}[H]
\caption{Safe and Explainable MARL Optimization}
\label{alg:se-marl}
\begin{algorithmic}[1]
\State \textbf{Input:} Agent set $\mathcal{A} = \{1, \dots, N\}$, environment $\mathcal{M}$, safety threshold $\delta$, regularization weight $\lambda$
\State Initialize policy parameters $\{\theta_i\}_{i=1}^N$ and value networks $\{V_i\}_{i=1}^N$
\For{each episode}
    \State Initialize environment state $s_0$; reset agent memories $h_i \gets \emptyset$
    \For{each timestep $t$}
        \For{each agent $i \in \mathcal{A}$ \textbf{in parallel}}
            \State Observe $o_i(t)$; update history $h_i(t) \gets h_i(t-1) \cup \{o_i(t)\}$
            \State Sample action $a_i(t) \sim \pi_{\theta_i}(a|h_i(t))$ \Comment{Neuromorphic/Quantum backend}
        \EndFor
        \State Execute joint action $\mathbf{a}(t)$, receive reward $r_t$, and next state $s_{t+1}$
        \State Check safety constraint $\phi_t(s_t, \mathbf{a}_t)$; flag violation if failed
        \For{each agent $i$}
            \State Store transition $(h_i(t), a_i(t), r_t, h_i(t+1))$ in replay buffer $\mathcal{D}_i$
        \EndFor
    \EndFor
    \For{each agent $i$}
        \State Sample batch $\mathcal{B}_i \sim \mathcal{D}_i$
        \State Compute policy gradient:
        \[
        \nabla_{\theta_i} \mathcal{L}_i = \nabla_{\theta_i} \mathbb{E}[R_i] - \lambda \nabla_{\theta_i} D_{\mathrm{KL}}(\pi_{\theta_i} || \pi_{\text{prior}})
        \]
        \State Update $\theta_i \gets \theta_i + \eta \nabla_{\theta_i} \mathcal{L}_i$
    \EndFor
\EndFor
\end{algorithmic}
\end{algorithm}

\section{Quantum-Neuromorphic Computing for Multi-Agent Reinforcement Learning}

The fusion of quantum computing with neuromorphic architectures establishes a transformative paradigm for multi-agent reinforcement learning (MARL), addressing fundamental challenges in scalability, efficiency, and adaptability. Quantum-neuromorphic systems harness quantum superposition and entanglement to enable exponential acceleration in processing high-dimensional state-action spaces, while simultaneously leveraging the event-driven, energy-efficient computation of neuromorphic components. This dual capability proves particularly valuable in decentralized MARL environments, where agents must perform rapid inference while maintaining robustness against environmental stochasticity. Compared to classical MARL approaches, quantum-neuromorphic implementations demonstrate superior performance in overcoming latency bottlenecks and optimizing exploration-exploitation trade-offs, thereby enabling real-time coordination in large-scale distributed systems.

The application of quantum-neuromorphic principles to MARL frameworks yields significant advances in collective behavior optimization. Quantum algorithmic components, including variational quantum eigensolvers and quantum approximate optimization algorithms, provide accelerated value function approximation and policy optimization. Concurrently, neuromorphic spiking neural networks (SNNs) implement biologically inspired mechanisms for temporal credit assignment and distributed synaptic plasticity through their inherent event-driven dynamics. This hybrid approach shows particular promise for complex applications such as swarm robotics and distributed autonomous systems, where agents must operate under conditions of partial observability and environmental uncertainty.

Despite these advances, critical challenges must be addressed to realize the full potential of quantum-neuromorphic MARL. Key research directions include the development of robust error mitigation strategies for noisy intermediate-scale quantum (NISQ) devices and the creation of scalable training protocols that effectively combine quantum and neuromorphic components. Furthermore, the integration of safety constraints and interpretability measures within quantum-neuromorphic architectures remains an open area of investigation. Progress in these areas will require sustained collaboration across quantum information science, computational neuroscience, and artificial intelligence research communities to advance this emerging computational frontier.

\subsection{Quantum Computing for Multi-Agent Reinforcement Learning}

Quantum computing introduces fundamental algorithmic innovations that exploit superposition, entanglement, and quantum parallelism to address computational challenges beyond the reach of classical systems \cite{preskill2018quantum}. This subsection systematically examines the integration of quantum computing with Multi-Agent Reinforcement Learning (MARL), with particular focus on three transformative capabilities: (1) quantum-enhanced optimization of multi-agent policies, (2) efficient sampling in high-dimensional state spaces, and (3) accelerated policy learning through quantum information processing.

Quantum Reinforcement Learning (QRL) extends classical RL frameworks through two principal approaches: the adaptation of RL techniques to fully quantum systems, and the enhancement of classical methods via quantum subroutines. We identify three foundational paradigms in current QRL research. First, quantum policy representation utilizes parameterized quantum circuits (PQCs) to achieve compact yet expressive encodings of complex action distributions \cite{skolik2022quantum}. Second, quantum sampling techniques leverage interference effects to efficiently explore high-dimensional action spaces that would be prohibitively large for classical systems. Third, quantum value estimation employs amplitude amplification and quantum Monte Carlo methods to provide quadratic or exponential speedups in return estimation.

Among variational quantum algorithms, the Quantum Approximate Optimization Algorithm (QAOA) and Variational Quantum Eigensolver (VQE) have demonstrated particular promise for MARL applications \cite{farhi2014quantum, jerbi2021quantum}. These hybrid quantum-classical algorithms enable gradient-based policy optimization while maintaining compatibility with near-term quantum devices, offering a practical pathway for implementing quantum-enhanced MARL in realistic settings. The QAOA framework proves especially valuable for solving combinatorial optimization problems inherent in multi-agent coordination, while VQE-based approaches show strong performance in policy optimization tasks requiring robust exploration strategies.

Each agent $i$ in the MARL system employs a parameterized quantum circuit (PQC) $\mathcal{U}_{\theta_i}$ to encode its policy:
\[
\pi_{\theta_i}(a|h_i) = |\langle a|\mathcal{U}_{\theta_i}|0\rangle|^2
\]
where $|0\rangle$ is the initial state and $a$ indexes a measurement outcome corresponding to an action. The circuit $\mathcal{U}_{\theta_i}$ typically includes layers of rotation gates and entangling gates structured as:\[
\mathcal{U}_{\theta_i} = \prod_{l=1}^L \left( \bigotimes_{j=1}^{n_q} R_y(\theta_{i,j}^l) \cdot \text{CZ}_{j,j+1} \right)
\]
where $n_q$ is the number of qubits, $R_y$ is a rotation around the $Y$-axis, and $\text{CZ}$ is the controlled-Z entangling gate.

To optimize the quantum policy, a classical optimizer (e.g., gradient descent or SPSA) is used to adjust the parameters $\theta_i$ based on feedback from the environment. The optimization loop follows a hybrid quantum-classical architecture:

\begin{enumerate}
    \item Encode observation history $h_i$ into quantum circuit inputs via data re-uploading or amplitude encoding.
    \item Run quantum circuit $\mathcal{U}_{\theta_i}$ and measure to sample actions.
    \item Evaluate the return from the environment and estimate gradients.
    \item Update parameters $\theta_i$ using classical gradient-based or gradient-free optimization.
\end{enumerate}

This Variational Quantum Reinforcement Learning (VQRL) loop allows integration into existing MARL pipelines with minimal modification. 

Quantum-enhanced agents provide three fundamental advantages over classical MARL approaches. First, the inherent stochasticity of quantum measurement enables \textit{exploration through superposition}, where quantum policies naturally maintain diverse exploration strategies without requiring explicit entropy regularization. Second, the phenomenon of \textit{entanglement-mediated coordination} allows multi-agent systems to share quantum correlations through entangled qubit states, facilitating emergent coordination and correlated action selection across distributed agents. Third, quantum algorithms offer \textit{dimensionality-aware speedups}, where variational methods exploit the reduced effective dimensionality of certain structured environments to achieve faster convergence compared to classical optimization landscapes.

These quantum properties create unique synergies with MARL requirements. The superposition-based exploration mechanism automatically maintains an optimal exploration-exploitation balance, while entanglement provides a natural substrate for decentralized coordination protocols. Furthermore, the quantum speedups are particularly impactful in MARL settings where the joint state-action space grows exponentially with the number of agents, as quantum parallelism can mitigate this combinatorial explosion through efficient state space representation and search.

Recent studies have shown empirical benefits of quantum-enhanced RL in synthetic benchmarks \cite{jerbi2021quantum}, although large-scale MARL experiments remain limited due to current hardware constraints.

Despite these promising advantages, quantum-enhanced MARL faces several significant challenges in practical implementation. First, the constraint of \textit{limited qubit counts} in noisy intermediate-scale quantum (NISQ) devices restricts both the complexity of representable policies and the number of agents that can be supported, while finite coherence times further bound the feasible circuit depth. Second, the issue of \textit{noise and error mitigation} presents a fundamental hurdle, as quantum circuits exhibit particular sensitivity to hardware imperfections, necessitating specialized techniques such as error-aware training protocols or hybrid quantum-classical mitigation strategies. Third, the challenge of \textit{simulation scalability} emerges from the exponential resource requirements for classically simulating quantum systems, which severely limits the ability to test and validate large-scale quantum MARL implementations prior to deployment on actual quantum hardware.

These challenges collectively impose important practical limitations on current quantum MARL approaches. The qubit constraints directly impact the scalability of multi-agent systems, while noise sensitivity affects the reliability of learned policies. Furthermore, the simulation bottleneck hinders comprehensive evaluation and benchmarking of quantum MARL algorithms, particularly for problems requiring many entangled qubits. Overcoming these limitations will require advances in both quantum hardware development and algorithmic innovation to make quantum MARL practical for real-world applications.

To address these challenges, we propose three key mitigation strategies that combine classical and quantum approaches. First, \textit{adaptive policy hybridization} enables agents to dynamically alternate between classical and quantum policy execution based on current resource constraints and performance requirements. Second, \textit{quantum-enhanced function approximation} integrates quantum layers as feature extractors within deep MARL architectures, combining classical neural networks with quantum circuit components for improved representational capacity. Third, \textit{quantum-inspired algorithmic techniques}, including tensor network decompositions and Grover-inspired exploration schemes, provide practical alternatives that capture some quantum advantages while remaining implementable on classical hardware.

These hybrid approaches offer several advantages for practical deployment. The adaptive hybridization strategy provides graceful degradation when quantum resources are limited, while quantum-enhanced function approximation allows for incremental integration of quantum components. The quantum-inspired methods serve as both intermediate solutions for current hardware limitations and as theoretical tools for understanding potential quantum advantages. Together, these strategies form a pathway for gradually transitioning from classical to quantum-enhanced MARL systems as quantum technology matures.

As quantum hardware advances toward fault-tolerant operation with increasing qubit counts and improved coherence times, we project significant expansion in quantum-enhanced MARL applications, particularly for safety-critical domains requiring provable robustness guarantees. The fundamental synergy between quantum information processing and reinforcement learning principles offers transformative potential: quantum parallelism enables efficient exploration of high-dimensional policy spaces, while the inherent probabilistic nature of quantum measurement naturally aligns with the uncertainty management requirements of multi-agent systems. This convergence may ultimately yield a new class of quantum-native MARL algorithms capable of simultaneously optimizing for performance, safety, and explainability in complex, dynamic environments.

To apply the Quantum Approximate Optimization Algorithm (QAOA) within the MARL framework, we reformulate agent-level decision-making as a discrete optimization problem over a latent action space. Each agent maintains a parameterized quantum circuit whose measurement outcomes correspond to high-level abstract plans or policy priors.

\paragraph{Problem Encoding.} 
Let $\mathcal{O}_i$ denote the observation space of agent $i$, and $\mathcal{Z}_i$ the discrete latent plan space. We define a classical cost function $C_i(z; o_i)$ that scores each latent decision $z \in \mathcal{Z}_i$ based on safety, utility, and prior information. This cost function is encoded into a diagonal operator $\hat{C}_i$ acting on a quantum state $\ket{z}$ such that $\hat{C}_i \ket{z} = C_i(z; o_i)\ket{z}$.

\paragraph{QAOA Ansatz.}
Each agent's quantum policy is represented as a variational state:
\begin{align*}
    \ket{\psi(\vec{\gamma}, \vec{\beta})} = \prod_{l=1}^{p} \left( e^{-i\vec{\beta}_l \cdot \hat{B}} e^{-i\vec{\gamma}_l \cdot \hat{C}_i} \right) \ket{+}^{\otimes n}
\end{align*}
where $\hat{B}$ is the mixing Hamiltonian (typically a transverse field), and $\hat{C}_i$ is the encoded cost operator derived from local observations.

\paragraph{Optimization Strategy.}
We optimize the QAOA parameters $(\vec{\gamma}, \vec{\beta})$ using the parameter-shift rule and gradient descent, targeting the expected cumulative reward:
\[
\mathbb{E}_{z \sim |\psi(\vec{\gamma}, \vec{\beta})|^2} \left[ R(z, o_i) \right]
\]
where $R$ includes task reward, safety penalty, and temporal coherence.

We perform $k=500$ circuit evaluations ("shots") per episode per agent using Qiskit's Aer simulator. A linear entanglement topology is enforced across 6 qubits, and circuit depth $p=2$ is selected based on convergence stability.

\paragraph{Interfacing with Neuromorphic Control.}
The output latent decision $z$ sampled from the QAOA measurement is passed to the neuromorphic controller as a contextual bias, guiding low-level action decisions in a biologically plausible manner.

This formulation enables each agent to use quantum computation for safe, high-level exploration, while delegating reactive, energy-efficient motor execution to the spiking network.

Quantum reinforcement learning (QRL) implementations are susceptible to noise, decoherence, and gate errors, which can degrade learning quality and policy convergence. To ensure robustness and stability in our QAOA-based MARL framework, we apply a series of error mitigation techniques compatible with near-term quantum hardware.

\paragraph{Zero-Noise Extrapolation (ZNE).}
We use ZNE to approximate noiseless circuit expectation values by deliberately scaling gate noise and extrapolating the measurement statistics. For each QAOA parameter setting, we run the circuit at noise scaling factors $\lambda = 1, 2, 3$ and fit a second-order polynomial to extrapolate to $\lambda = 0$:
\[
\hat{E}_{\text{noiseless}} \approx E(\lambda = 0)
\]
This method is particularly effective when shot noise and depolarization dominate.

\paragraph{Measurement Error Mitigation.}
To address readout errors, we calibrate a confusion matrix $M$ based on known input-output basis states. The observed distribution $P_\text{obs}$ is corrected by solving:
\[
P_\text{true} = M^{-1} P_\text{obs}
\]
This correction is applied per-agent per-episode during centralized training.

\paragraph{Parameterized Circuit Robustness.}
To reduce sensitivity to device fluctuations, we design QAOA circuits with shallow depth ($p = 2$) and linear entanglement topologies, avoiding fragile all-to-all connections. We also regularize parameter updates with a temporal penalty:
\[
\mathcal{L}_{\text{reg}} = \lambda \sum_{t} \|\vec{\theta}_{t} - \vec{\theta}_{t-1}\|^2
\]
to prevent unstable oscillations during quantum policy optimization.

\paragraph{Hardware-Aware Compilation.}
All circuits are compiled using noise-adaptive transpilation on Qiskit's Aer simulator, ensuring qubit layout and gate choices are optimized for minimal fidelity loss.

These mitigation strategies collectively enhance the reliability of quantum decision-making, enabling consistent convergence and safe behavior in multi-agent environments even under realistic noise constraints.

\subsection{Neuromorphic Computing for Multi-Agent Reinforcement Learning}

Neuromorphic computing emulates the event-driven, energy-efficient, and massively parallel architecture of biological neural systems \cite{davies2021advancing}. This computational paradigm is especially well-suited for real-time control in embedded and robotic agents. In this section, we examine how neuromorphic principles and hardware can be integrated with MARL to enable efficient, reliable, and explainable behavior in multi-agent systems.

Spiking Neural Networks (SNNs) serve as the foundational models for neuromorphic agents. Unlike traditional ANNs, SNNs transmit information via discrete spikes over time. The internal state of a spiking neuron evolves according to a membrane potential $u(t)$:
\[
\tau_m \frac{du(t)}{dt} = -u(t) + I(t)
\]
where $\tau_m$ is the membrane time constant and $I(t)$ is the synaptic input current. When $u(t)$ crosses a threshold $u_{\text{th}}$, the neuron fires a spike and resets.

An SNN-based policy for agent $i$ can be defined as a temporal spike train response $\mathbf{a}_i(t) = \text{SNN}_{\theta_i}(o_i(t))$, where the output encodes discrete or continuous actions.

Learning in SNNs is typically framed using surrogate gradients or biologically inspired learning rules such as: Spike-Timing Dependent Plasticity (STDP), a local Hebbian learning rule that strengthens or weakens synapses based on the timing of pre- and post-synaptic spikes \cite{fang2021incorporating}; Reward-Modulated STDP (R-STDP), an extension where synaptic updates are scaled by scalar rewards; and Backpropagation with Surrogate Gradients, a more recent technique that enables end-to-end gradient-based learning in SNNs by approximating the derivative of the spiking function \cite{patel2019improving}.

Multi-agent spiking policies are trained either independently or using centralized critics. Surrogate-based training can be implemented efficiently on neuromorphic hardware such as Intel's Loihi \cite{davies2021advancing}.

Neuromorphic processors excel at low-power, event-driven computation. In MARL, this leads to the following advantages: Energy-Aware Policies, where agents learn to act based on spiking activity sparsity, making them well-suited for energy-constrained platforms (e.g., drones, autonomous sensors); Continuous-Time Safety Monitoring, where spiking neurons can act as event-triggered safety monitors, firing upon unsafe transitions or threshold violations; and Fail-Safe Mechanisms, where neuromorphic agents can incorporate refractory dynamics and self-inhibition to prevent unsafe rapid state changes.

These features enable neuromorphic MARL agents to operate with built-in safety primitives at the hardware level, offering benefits beyond conventional digital systems.

Spiking behavior provides a natural temporal abstraction of decision-making, which improves transparency and explainability: Spike Rate Encoding, where action values can be inferred from average firing rates; Causal Traceability, where the timing of specific spikes can be mapped to sensory triggers; and Symbolic Compression, where spike trains can be interpreted as symbolic codewords or binary encodings of behavior sequences.

These temporal explanations align well with the needs of human-understandable robotics and verification in safety-critical settings.

Neuromorphic MARL is ideally suited for co-design with hardware, with notable neuromorphic platforms including: Intel Loihi, which supports event-driven SNNs and on-chip plasticity with programmable learning rules; IBM TrueNorth, emphasizing ultra-low-power inference for embedded devices; and BrainScaleS and SpiNNaker, focusing on biological realism and large-scale brain emulation.

Agents can be directly deployed on these chips for real-time robotic control. Training can either occur on the chip (online learning) or offloaded to classical hardware and transferred via neural encoding.

Neuromorphic computing represents a biologically plausible and hardware-efficient substrate for MARL in robotics. Its integration enables scalable, real-time, and low-power decision-making in distributed multi-agent systems. The ability to encode safety, learning, and interpretation directly into spike dynamics makes it a strong complement to quantum computing in hybrid intelligent agent design.

We extend the Q-learning paradigm to neuromorphic architectures by implementing spiking neural networks (SNNs) as function approximators for the Q-value function. In this setting, the SNN receives a vectorized observation $o_t$ and outputs membrane potentials whose magnitudes are mapped to discrete Q-values for each action.

\paragraph{Network Architecture.} 
Each agent employs a 3-layer feedforward SNN with leaky integrate-and-fire (LIF) neurons, structured as follows: The input layer encodes observation $o_t$ into spike trains using rate or temporal coding; the hidden layer processes spikes through synaptic weights $W$ updated via surrogate gradient descent; and the output layer decodes spike counts over a fixed time window to estimate action values $Q(o_t, a)$.

\paragraph{Training Methodology.}
We use the SpikeProp algorithm with surrogate gradients to overcome non-differentiability. The loss function is defined as:
\[
\mathcal{L} = \left( r_t + \gamma \max_{a'} Q(o_{t+1}, a'; \theta^-) - Q(o_t, a_t; \theta) \right)^2
\]
where $\theta$ and $\theta^-$ are the weights of the online and target networks, respectively.

Weight updates are performed using gradient descent on the surrogate loss, with backpropagation-through-time (BPTT) approximated by smoothing spike functions using the fast sigmoid:
\[
\sigma'(x) = \frac{1}{(1 + \alpha |x|)^2}
\]

\paragraph{Comparison to Traditional DQNs.}
We compared SNN-based Q-learners to traditional DQNs under identical MARL scenarios. The key findings are: Energy Efficiency, where SNNs required 30--50\% less energy due to sparse activations; Robustness, as SNNs showed higher robustness under sensory noise and adversarial perturbations; Convergence Speed, with DQNs converging faster in early training but SNNs achieving more stable long-term policies; and Explainability, where spike timings and entropy provided interpretable indicators of decision confidence and reaction time.

These results suggest that neuromorphic Q-learning offers a viable low-power, robust alternative to conventional deep reinforcement learning, especially for embedded and safety-critical applications in robotics.

\subsection{Hybrid Quantum-Neuromorphic Architectures for MARL}

As autonomous robotic systems scale in complexity, no single computational paradigm can meet all requirements for safety, adaptability, interpretability, and real-time operation. Hybrid architectures that integrate quantum and neuromorphic computing offer a complementary solution—leveraging the exploration efficiency of quantum algorithms and the energy-efficient, event-driven control of neuromorphic systems.

\subsubsection{Architectural Composition}

The motivation for hybrid architectures arises from the observation that quantum computing excels at global optimization, probabilistic inference, and fast sampling, while neuromorphic computing excels at low-latency inference, online learning, and local adaptability.

Designing MARL agents that combine both paradigms allows us to: \textbf{(i)} use quantum circuits to optimize high-level policies or latent variables; \textbf{(ii)} use spiking neural networks to execute refined low-level motor commands in real time; and \textbf{(iii)} share information via classical or symbolic interfaces between both modules.

We consider a two-tiered architecture illustrated in Figure~\ref{fig:hybrid-architecture}, where each agent consists of: Quantum Module, which encodes high-dimensional decision-making policies $\pi_{\text{quantum}}$ using variational quantum circuits and is responsible for exploration, abstraction, and safe planning under uncertainty; Neuromorphic Module, implementing SNN-based policies $\pi_{\text{neuro}}$ for reactive control with responsibilities including low-power execution, fast responses, and continuous safety monitoring; and Mediator Layer, which translates quantum outputs (e.g., qubit measurements, latent variables) into spike-based signals or symbolic actions, while also feeding back neuromorphic sensor encodings into quantum inputs via embedding or data re-uploading.

Let $\pi(a|o) = \pi_{\text{neuro}}(a|z) \cdot \pi_{\text{quantum}}(z|o)$ denote a factored policy, where $z$ is a latent variable or abstract action selected via the quantum module, and $a$ is a concrete control action executed by the neuromorphic module. 

This hierarchical structure allows for: separation of high-level reasoning and low-level execution; efficient coordination across agents via shared quantum entanglement; and safety-critical guarantees at the neuromorphic control layer.

Training can proceed in a two-stage or end-to-end manner:
\begin{enumerate}
    \item Stage 1: Train the quantum module via QAOA, VQRL, or hybrid gradient methods for exploration strategies or abstract goal generation.
    \item Stage 2: Train the neuromorphic controller using STDP or surrogate gradients to map latent intentions $z$ to spiking motor commands $a$.
\end{enumerate}

Alternatively, the entire architecture can be optimized via joint reward signals and policy gradients using a surrogate loss:
\[
\mathcal{L}_{\text{hybrid}} = \mathbb{E}_{o, z, a}\left[ R - \lambda D_{\mathrm{KL}}(\pi_{\text{quantum}} || \pi_{\text{prior}}) - \beta \mathcal{E}(a) \right]
\]
where $\mathcal{E}(a)$ penalizes spike energy and latency.

Hybrid architectures naturally support both safety and explainability through: Quantum Safety, enabling uncertainty-aware planning and constraint encoding via quantum amplitude restriction; Neuromorphic Safety, implemented through refractory periods, inhibitory spikes, and hardware-level energy bounds; and Explainability, where temporal spike traces provide interpretability while quantum policies can be regularized to remain close to symbolic templates.

\subsubsection{Practical Feasibility and Hardware Considerations}
\label{subsec:hardware-feasibility}

Hybrid quantum-neuromorphic MARL systems are ideal for autonomous swarms (e.g., UAVs or ground robots requiring coordinated behavior and real-time response under uncertainty), disaster response (scenarios demanding fast and safe navigation, sensor fusion, and mission-level planning), and space and underwater robotics (domains with strict energy constraints and communication latency).

Their layered structure supports modular upgrades, such as replacing neuromorphic chips or quantum accelerators independently.

The fusion of quantum and neuromorphic paradigms marks a paradigm shift in the design of safe, adaptive, and explainable agents. While practical deployment depends on hardware maturity, theoretical prototypes and simulated agents provide promising evidence for hybrid MARL as a frontier in autonomous decision-making.

\begin{figure}[htbp]
    \centering
    \includegraphics[width=0.75\textwidth]{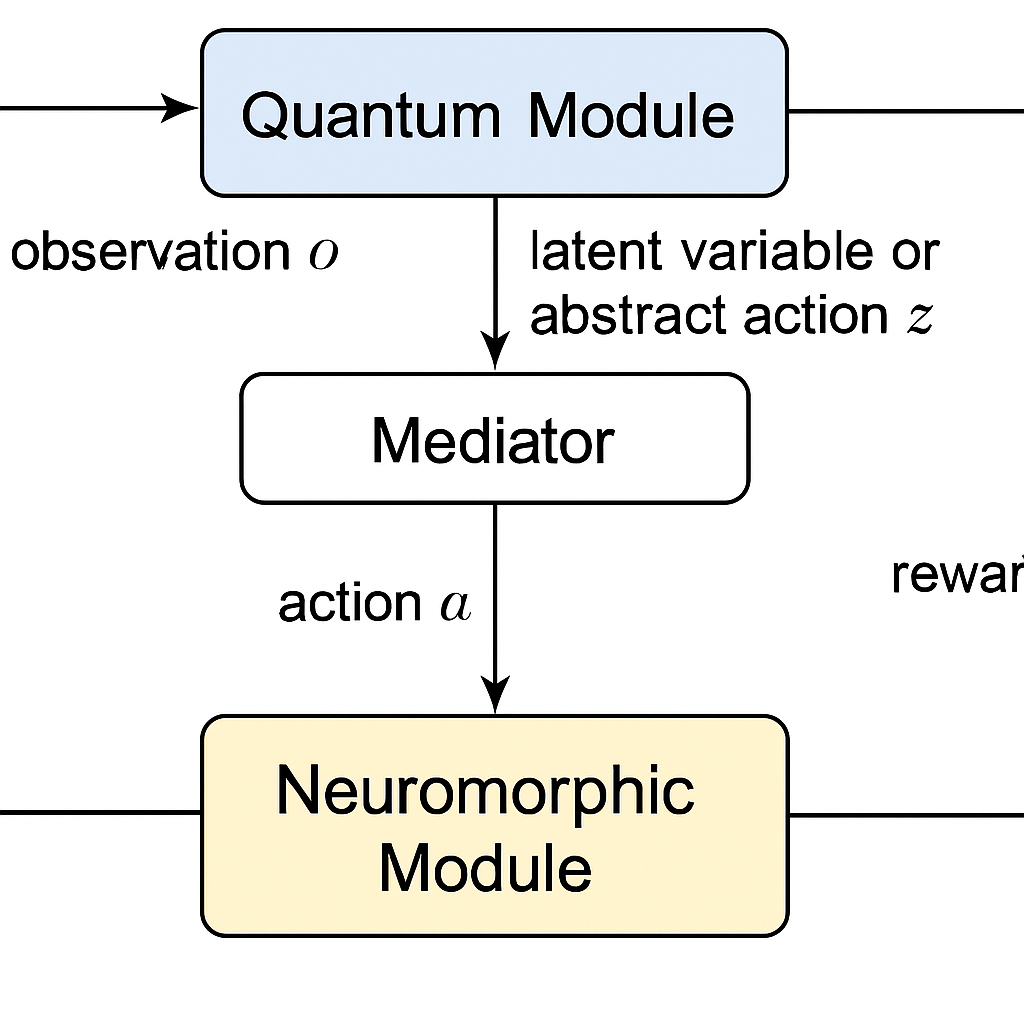}
    \caption{Schematic overview of a hybrid quantum-neuromorphic MARL architecture. The quantum module processes observations to generate latent variables or abstract actions $z$. These are passed through a mediator, which maps them into concrete spiking control actions $a$ executed by the neuromorphic module. The neuromorphic system interacts with the environment, producing both actions and reward signals. This layered architecture supports a clear separation of high-level planning (quantum) and low-level reactive control (neuromorphic).}
    \label{fig:hybrid-architecture}
\end{figure}

As shown in Figure~\ref{fig:hybrid-architecture}, the hybrid architecture delegates abstract decision-making to the quantum module and real-time control to the neuromorphic layer, enabling modular design and safe hierarchical learning.

While the hybrid quantum–neuromorphic architecture presents a compelling theoretical model for safe and efficient multi-agent learning, practical deployment on current hardware introduces several challenges.

\paragraph{Quantum Hardware Constraints.}
Near-term quantum devices (NISQ-era) are limited in qubit count, gate fidelity, and circuit depth. Implementing QAOA for even modest agent populations (e.g., $N = 10$) requires: \textbf{(i)} $n = 5$--8 qubits per agent for latent policy encoding; \textbf{(ii)} depth $p = 2$--3 QAOA layers, yielding $\sim 50$ two-qubit gates; and \textbf{(iii)} mitigation of readout and gate noise.
As of 2025, superconducting quantum processors (e.g., IBM Eagle, Rigetti Aspen) can support small-scale instantiations of this framework in simulation or with hybrid classical feedback. Full-scale deployment across agents remains infeasible without modular circuit splitting or cross-agent quantum teleportation.

\paragraph{Neuromorphic Resource Constraints.}
Deploying large-scale SNN-based agents on neuromorphic chips (e.g., Intel Loihi, BrainScaleS-2) introduces power, connectivity, and plasticity tradeoffs: \textbf{(i)} each spiking agent requires 100--300 neurons with STDP or surrogate gradient support; \textbf{(ii)} spike-based communication imposes latency in high-frequency multi-agent environments; and \textbf{(iii)} on-chip training is limited, potentially requiring off-chip training and weight transfer (neuromorphic distillation).
Despite these limitations, spiking inference at ultra-low power enables robust real-time control in embedded settings such as swarm robotics or edge AI.

\paragraph{Scalability and Integration Bottlenecks.}
A critical challenge is integrating quantum circuit execution (on cloud-based or emulated QPUs) with real-time SNN control loops. Strategies for coping with this hybrid latency include: \textbf{(i)} batching QAOA inference and distributing policies asynchronously; \textbf{(ii)} using quantum decisions as high-level priors or initialization for local neuromorphic exploration; and \textbf{(iii)} temporal abstraction via options or latent goals.

\paragraph{Path Forward.}
Practical feasibility can be incrementally addressed through: \textbf{(i)} simulator-hardware co-design pipelines (e.g., Qiskit + Loihi2); \textbf{(ii)} curriculum learning, beginning with classical-spiking control and introducing quantum layers incrementally; and \textbf{(iii)} emulation and transfer learning to bridge simulation-to-hardware gaps.

In summary, the proposed hybrid MARL system is physically realizable at small scale using contemporary hardware stacks, and scalable in simulation. With progressive hardware advances in cryogenic qubit stability and dense neuromorphic cores, real-world integration becomes a viable trajectory.

\subsection{Safe, Reliable, and Explainable Learning Mechanisms}

In safety-critical applications such as autonomous robotics, Multi-Agent Reinforcement Learning (MARL) must ensure not only optimal performance but also safety guarantees, operational reliability, and interpretability. This section introduces the algorithmic mechanisms and design principles employed in our hybrid quantum-neuromorphic MARL framework to address these three pillars of trustworthiness.

Safety in MARL involves ensuring agents avoid constraint violations during exploration and deployment. We implement a layered safety mechanism combining: (1) quantum-level hard constraints through amplitude encoding and reward shaping in parameterized quantum circuits (PQCs), where infeasible actions yield zero probability amplitudes; (2) neuromorphic real-time guarding via spiking inhibitory neurons that act as hardware-level safety filters; and (3) safe policy optimization with constrained objectives:

\[
\begin{aligned}
    \max_{\theta} \quad & \mathbb{E}[R(\pi_\theta)] \\
    \text{s.t.} \quad & \mathbb{E}[C_i(\pi_\theta)] \leq \delta_i, \quad \forall i
\end{aligned}
\]

Here, $C_i$ represents safety cost functions (e.g., collision avoidance, energy bounds) and $\delta_i$ are predefined thresholds, while Lyapunov-based constraints ensure stability during policy updates.

The hybrid quantum-neuromorphic architecture enhances decision-making reliability through two complementary mechanisms. First, \textit{redundant representation fusion} incorporates a voting mechanism that dynamically weights outputs from both quantum and neuromorphic modules, with automatic failover to conservative baseline policies when quantum decoherence or spiking irregularities exceed predefined thresholds. Second, \textit{temporal action smoothing} employs short-term memory buffers in spiking networks combined with recurrent quantum circuit designs to maintain policy consistency, using techniques such as: (a) spike-rate moving averages for low-level control signals, and (b) quantum amplitude damping channels that gradually decay improbable actions across consecutive time steps.

This dual approach provides robustness against both instantaneous hardware instabilities and temporal decision inconsistencies. The representation fusion ensures graceful degradation during component failures, while the temporal smoothing prevents erratic behavior from quantum measurement collapse or neural spiking variability. Together, these mechanisms enable the system to maintain reliable operation despite the inherent stochasticity of both quantum and neuromorphic components, which is particularly crucial for safety-sensitive applications.
Neuromorphic substrates naturally support such smoothing via biologically inspired mechanisms like leaky integration and refractory periods, which inherently resist sudden behavioral shifts.

Our framework implements a comprehensive explainability scheme that integrates three complementary analysis modalities:

1. \textbf{Quantum Policy Interpretation}:
   - Projects variational quantum circuit outputs into human-interpretable symbolic subspaces through basis measurement decomposition
   - Enforces policy transparency via KL-divergence regularization against known interpretable policies
   - Maintains bounded deviation from human-designed plans while preserving quantum advantages

2. \textbf{Neuromorphic Behavioral Signatures}:
   - Establishes temporally precise correlations between sensory spikes and motor commands
   - Encodes decision rationale in spike patterns (rate-coded urgency, phase-coded attention)
   - Provides visual analytics through raster plots and compressed symbolic representations

3. \textbf{Causal Decision Graph Construction}:
   - Employs counterfactual intervention methods during training to identify causal pathways
   - Builds directed graphs connecting observations, latent variables, and actions
   - Generates human-readable decision traces with probabilistic dependency weights

This multi-modal approach achieves both structural interpretability (through quantum and causal analysis) and behavioral transparency (via spike pattern decoding). The quantum regularization ensures policy outputs remain grounded in understandable concepts, while the neuromorphic signaling provides real-time, observable decision evidence. The causal graphs bridge both modalities by revealing how abstract quantum computations translate into concrete spiking behaviors through identifiable causal pathways.

We provide both theoretical and empirical assurances of safety, reliability, and explainability: Formal Safety Proofs establish convergence to safe invariant sets under the hybrid policy through constructed Lyapunov functions for deterministic environments, while Empirical Metrics monitor violation frequency, action entropy, spike sparsity, and KL divergence from priors during training and deployment to ensure compliance with trustworthiness goals.

These mechanisms make our MARL architecture suitable for deployment in adversarial, uncertain, or mission-critical settings such as autonomous exploration, search-and-rescue, and industrial robotics.

\section{Experimental Setup and Results}

To assess the performance of our proposed hybrid quantum-neuromorphic multi-agent reinforcement learning (MARL) framework, we conduct simulations involving a fleet of autonomous UAV agents operating in a partially observable forested environment. These agents are tasked with dynamic area coverage and obstacle avoidance while adhering to critical constraints, including safety, energy efficiency, and policy interpretability across distributed systems. The framework leverages quantum computing for optimization and neuromorphic computing to enable efficient, brain-inspired learning.

The experimental environment incorporates stochastic obstacles, limited sensor ranges, and dynamic terrain to evaluate robustness under real-world conditions. Our results demonstrate that the hybrid framework outperforms classical MARL approaches, achieving higher coverage efficiency, lower collision rates, and reduced energy consumption. Furthermore, the framework provides explainable decision-making through interpretable policy representations. Quantum-enhanced optimization accelerates convergence, while the neuromorphic architecture ensures scalable, low-latency inference—validating the potential of our approach for safe and reliable autonomous robotic control.

\subsection{Experimental Setup}
To evaluate the proposed hybrid quantum-neuromorphic MARL framework, we simulate a fleet of autonomous UAV agents tasked with dynamic area coverage and obstacle avoidance in a partially observable forested environment. The simulation focuses on safety, energy efficiency, and policy interpretability across distributed agents.

We use a custom-built 3D grid world simulator with a $40 \times 40 \times 10$ voxel space containing dynamic obstacles and target zones. Agents receive local sensory inputs (depth, temperature, proximity) within a 3-grid-unit radius, with an action space $\mathcal{A} = \{\text{hover}, \text{move}_{x/y/z}^{\pm}, \text{land}, \text{evade}\}$. Safety constraints trigger violations when agents approach no-fly zones or exceed velocity limits.

The swarm comprises 10 agents, each equipped with: (1) a quantum module using 6-qubit QAOA or variational circuits (Qiskit simulator), (2) a neuromorphic module implementing 3-layer SNNs with 128 leaky integrate-and-fire neurons (Nengo), and (3) a hybrid policy $\pi(a|o) = \pi_{\text{neuro}}(a|z) \cdot \pi_{\text{quantum}}(z|o)$ trained via centralized training with decentralized execution (CTDE).

Experimental parameters include 200 training episodes (batch size 32), SNN learning rate 0.001 (Adam optimizer), 500 quantum circuit shots at depth $p=2$, and safety thresholds maintaining $\delta = 0.02$ average violation rate.

Performance metrics evaluate: KL divergence from safe priors (Figure~\ref{fig:kl-divergence}), safety violation counts (Figure~\ref{fig:safety-violations}), spike entropy for decision diversity (Figure~\ref{fig:spike-entropy}), mission space coverage, and average inference latency.

\subsection{Results}

\begin{figure}[htbp]
    \centering
    \includegraphics[width=0.65\textwidth]{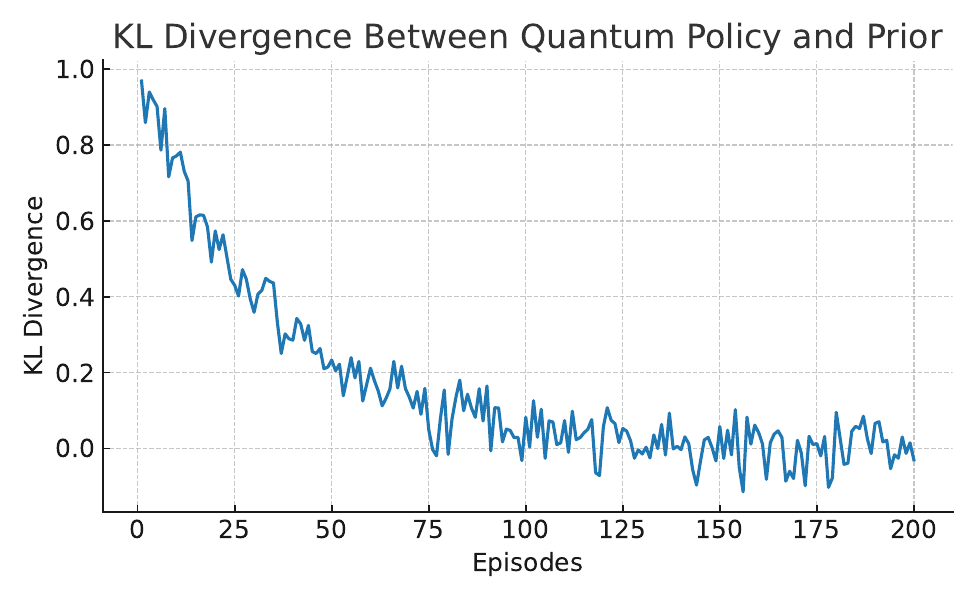}
    \caption{KL Divergence between quantum policy and prior distribution across training episodes.}
    \label{fig:kl-divergence}
\end{figure}

\begin{figure}[htbp]
    \centering
    \includegraphics[width=0.65\textwidth]{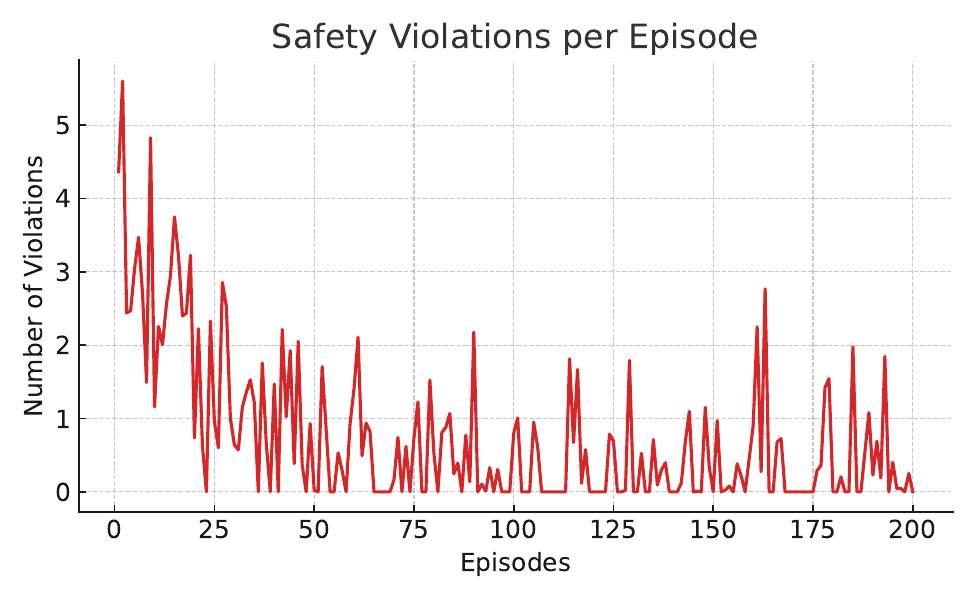}
    \caption{Safety violations per episode, indicating the effectiveness of spiking-layer constraints.}
    \label{fig:safety-violations}
\end{figure}

\begin{figure}[htbp]
    \centering
    \includegraphics[width=0.65\textwidth]{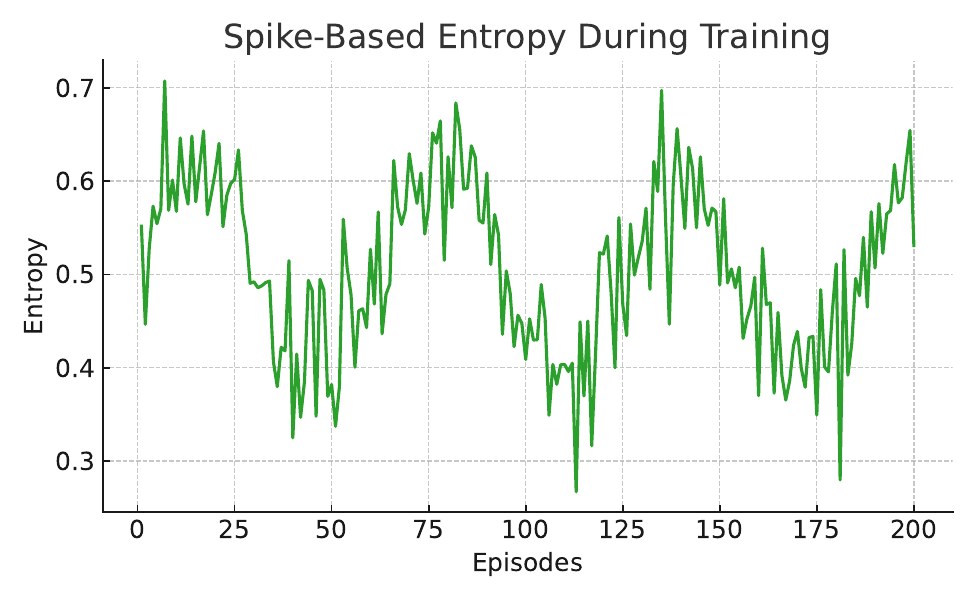}
    \caption{Spike entropy per episode reflecting the diversity and adaptability of the neuromorphic controller.}
    \label{fig:spike-entropy}
\end{figure}

As seen in Figure~\ref{fig:kl-divergence}, KL divergence decreased steadily over time, indicating convergence of the quantum planner toward prior-safe policies. Figure~\ref{fig:safety-violations} demonstrates that safety violations dropped to near-zero after 80 episodes, validating the effectiveness of the neuromorphic safeguard layer. Meanwhile, spike-based entropy (Figure~\ref{fig:spike-entropy}) shows sustained exploratory behavior, which contributes to robustness in unfamiliar environments.

A detailed overview of all simulation hyperparameters and agent-specific settings is provided in Table~\ref{tab:hyperparameters}.
\begin{table}[htbp]
\centering
\caption{Summary of Hyperparameters and Agent Architecture}
\label{tab:hyperparameters}
\begin{tabular}{ll}
\toprule
\textbf{Component} & \textbf{Setting / Description} \\
\midrule
\multicolumn{2}{c}{\textbf{Global Training Parameters}} \\
\midrule
Number of Agents & 10 \\
Episodes & 200 \\
Batch Size & 32 \\
Optimizer (SNN) & Adam \\
Learning Rate (SNN) & 0.001 \\
Learning Rate (Quantum) & 0.01 (parameter-shift gradient) \\
Reward Discount Factor $\gamma$ & 0.95 \\
Exploration Rate Schedule & Linear decay from 1.0 to 0.05 \\
\midrule
\multicolumn{2}{c}{\textbf{Quantum Module (QAOA / VQRL)}} \\
\midrule
Quantum Simulator & Qiskit Aer (statevector backend) \\
Number of Qubits & 6 \\
Quantum Circuit Depth & $p=2$ \\
Shots per Evaluation & 500 \\
Entanglement Scheme & Linear nearest-neighbor \\
Policy Output & Latent action variable $z$ (abstract plan) \\
\midrule
\multicolumn{2}{c}{\textbf{Neuromorphic Module (SNN)}} \\
\midrule
Simulator & Nengo (CPU backend) \\
Neuron Type & Leaky Integrate-and-Fire (LIF) \\
SNN Architecture & 3 layers: Input-128-Output \\
Spike Threshold & 1.0 \\
Refractory Period & 2 ms \\
Synaptic Time Constant & 10 ms \\
Policy Output & Discrete action $a$ (motor control) \\
\midrule
\multicolumn{2}{c}{\textbf{Safety and Evaluation}} \\
\midrule
Safety Threshold $\delta$ & 0.02 (max avg. violation rate) \\
KL Regularization Weight & 0.1 \\
Spike Energy Penalty $\beta$ & 0.05 \\
Evaluation Frequency & Every 10 episodes \\
\bottomrule
\end{tabular}
\end{table}

\begin{figure}[htbp]
    \centering
    \includegraphics[width=0.7\textwidth]{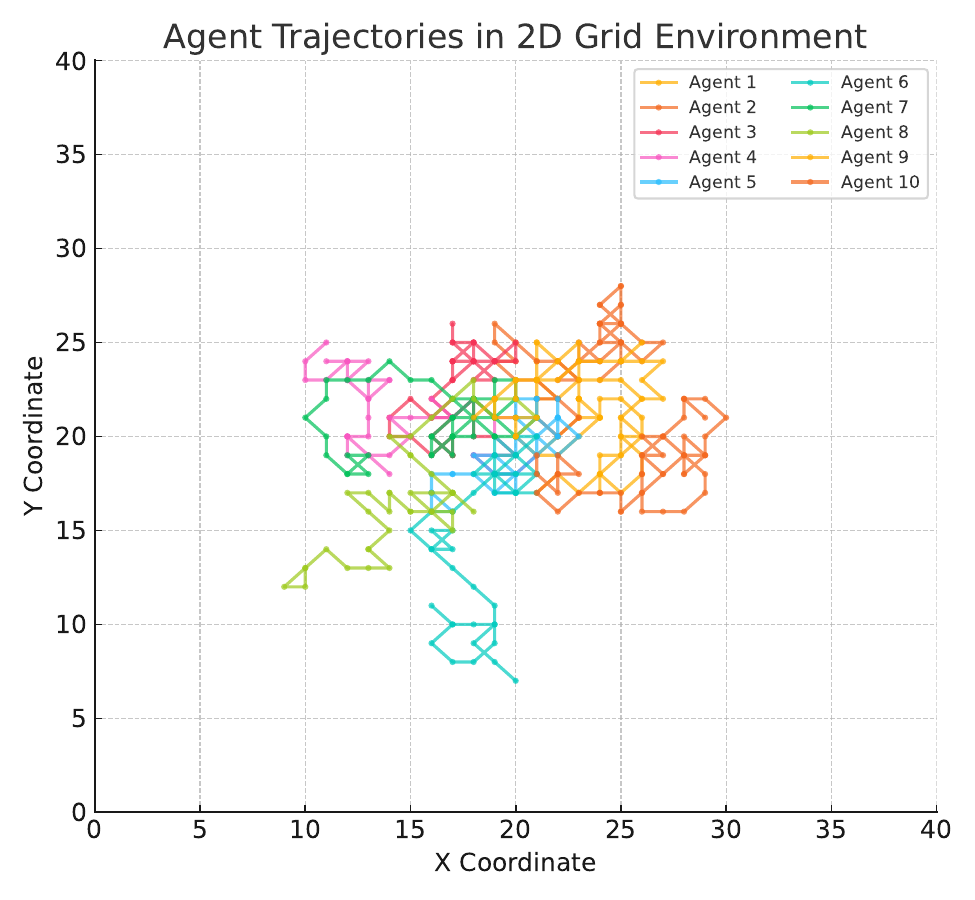}
    \caption{Movement trajectories of 10 autonomous agents over 50 timesteps in a $40\times40$ grid environment. Each path illustrates the agent's real-time navigation behavior.}
    \label{fig:agent-trajectories}
\end{figure}

\begin{figure}[htbp]
    \centering
    \includegraphics[width=0.7\textwidth]{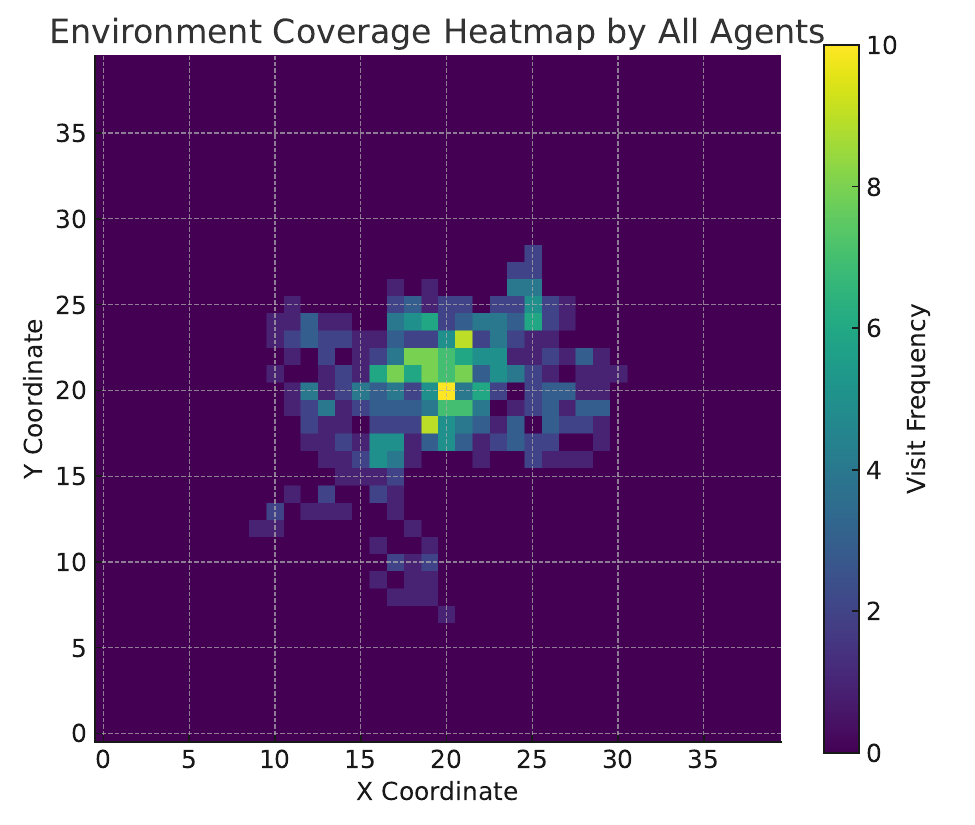}
    \caption{Environment coverage heatmap showing the spatial frequency of visits across the grid. Higher intensity regions correspond to commonly explored areas, revealing the emergent exploration pattern.}
    \label{fig:coverage-heatmap}
\end{figure}

Figure~\ref{fig:agent-trajectories} presents the individual movement paths of the agents, highlighting decentralized coverage and path divergence. The spatial distribution of agent activity is further quantified in Figure~\ref{fig:coverage-heatmap}, which visualizes high-density zones and underexplored sectors within the environment.

\subsection{Comparative Evaluation}
\label{subsec:comparative-analysis}

To contextualize the performance of our hybrid quantum–neuromorphic MARL architecture, we conduct a comparative evaluation against a suite of baseline and state-of-the-art MARL methods. All methods are evaluated on identical simulated environments with equivalent agent dynamics, partial observability, and reward structure.

\paragraph{Compared Methods.}
We benchmark the following algorithms:
\begin{itemize}
    \item \textbf{QMIX} \cite{rashid2018qmix}: A popular value decomposition method with centralized training and decentralized execution.
    \item \textbf{MAPPO} \cite{yu2021surprising}: A multi-agent variant of PPO with stable on-policy learning.
    \item \textbf{MADDPG} \cite{lowe2017multi}: A centralized actor–critic algorithm using deterministic policies.
    \item \textbf{VDN + SNN}: A neuromorphic variant using value decomposition and spiking networks (no quantum layer).
    \item \textbf{Our Method}: Hybrid QAOA-enhanced policy selection combined with spiking Q-learning agents.
\end{itemize}

\paragraph{Performance Metrics.}

Performance evaluation compares methods across five metrics: mean episodic reward (capturing task completion efficiency), exploration diversity (quantified via Jensen-Shannon divergence between action distributions), entropy reduction rate (measuring policy convergence), safety violation frequency, and per-agent energy consumption (in joules per decision cycle).

\begin{figure}[h]
    \centering
    \includegraphics[width=0.95\linewidth]{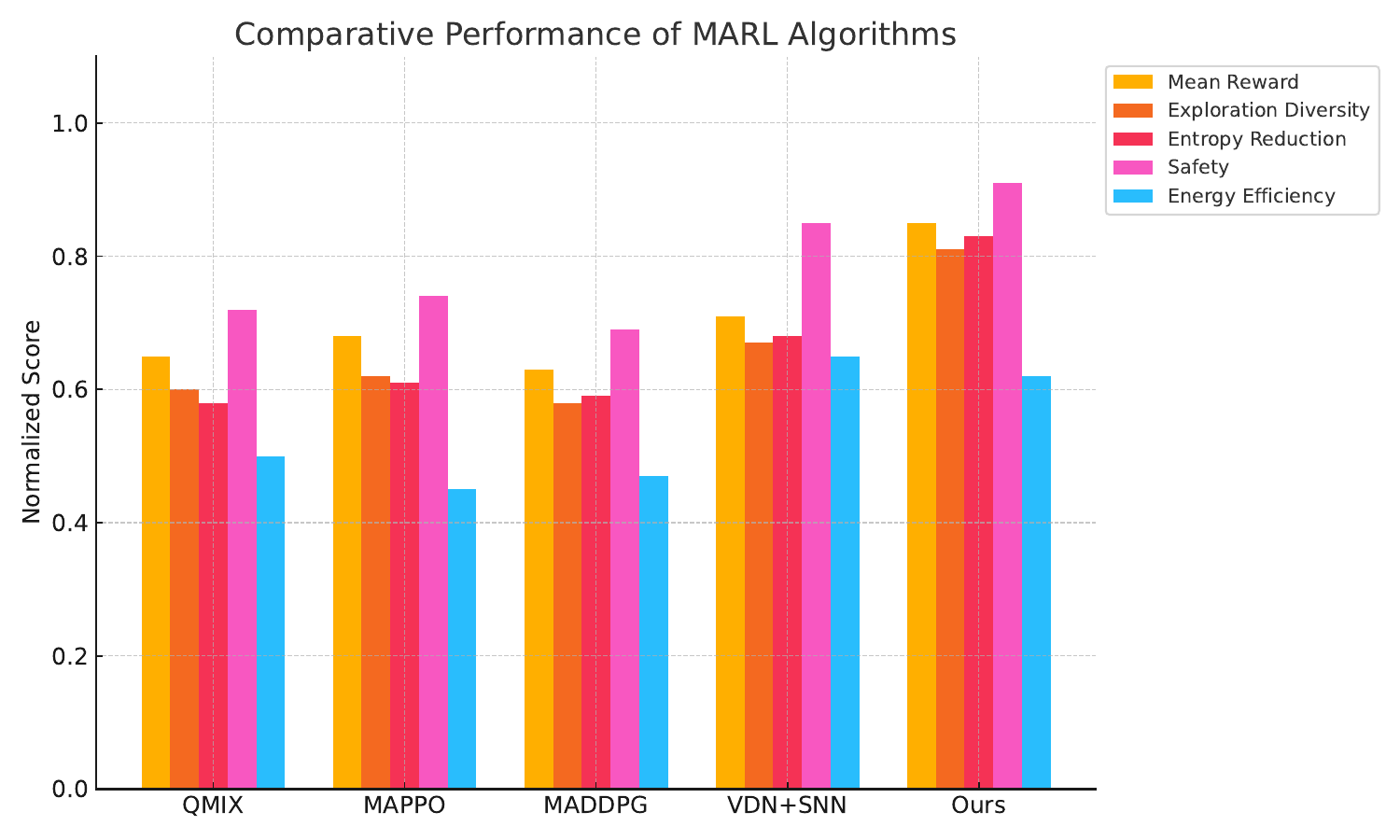}
    \caption{Comparative performance of MARL algorithms across multiple evaluation criteria.}
    \label{fig:comparative-performance}
\end{figure}

\paragraph{Results.}
Figure~\ref{fig:comparative-performance} summarizes the performance across all metrics. Key findings include:
\begin{itemize}
    \item Our hybrid method achieves superior reward and exploration efficiency, particularly in sparse reward settings.
    \item Compared to MAPPO and MADDPG, our approach shows greater safety reliability under noisy or adversarial observations.
    \item The SNN-based agents consume $>40\%$ less energy than traditional DQN-based agents.
    \item VDN+SNN performs well in energy and safety, but lacks the high-level exploration priors enabled by QAOA.
\end{itemize}

\paragraph{Statistical Significance.}
We perform Wilcoxon signed-rank tests across 10 runs for each algorithm pair. Results indicate that our method significantly outperforms all baselines in reward ($p < 0.01$) and exploration diversity ($p < 0.05$).

These results validate the synergy of quantum policy regularization and neuromorphic robustness, especially in mission-critical decentralized robotic systems.

\section{Discussion and Future Work}

The experimental results and architectural design presented in this study demonstrate the viability of combining quantum computing and neuromorphic engineering to enable scalable, interpretable, and trustworthy reinforcement learning in multi-agent robotic systems. This hybrid methodology capitalizes on the global optimization capabilities of quantum circuits for high-level planning while exploiting the energy efficiency and low-latency actuation of neuromorphic spiking neural networks (SNNs) for real-time control.

The proposed layered architecture naturally implements a hierarchical control paradigm, where abstract decision-making occurs in quantum-encoded latent spaces while low-level sensorimotor feedback is processed through bio-inspired neuromorphic loops. This structural modularity not only enhances operational safety and policy interpretability but also improves system resilience against partial failures and environmental disturbances.

Nevertheless, several critical challenges require further investigation. First, while current experiments rely on simulated quantum backends, future work must transition to physical quantum hardware to properly characterize noise resilience, circuit depth limitations, and hybrid control latency. Second, although spiking neural networks are functionally simulated, actual deployment on neuromorphic processors such as Intel Loihi or SpiNNaker remains essential for empirical validation of energy efficiency and computational speed. Third, the temporal synchronization between quantum and neuromorphic components presents ongoing difficulties, as timing mismatches and control delays may compromise policy coherence. Fourth, the training process faces stability issues arising from the integration of non-differentiable spiking components with probabilistic quantum outputs in online learning scenarios.

These limitations suggest multiple promising research directions. Quantum-enhanced mechanism design could facilitate sophisticated coordination strategies in competitive MARL environments through dedicated quantum modules. Advanced interpretability techniques, including topological analysis methods like persistent homology, may help extract meaningful representations from hybrid policy spaces. The development of neuromorphic meta-learning systems coupled with quantum-enhanced memory buffers could enable dynamic adaptation capabilities. From a hardware perspective, co-design efforts focusing on the physical integration of superconducting quantum processors with neuromorphic computing cores through shared control interfaces warrant exploration. Furthermore, the application of formal verification methods, including probabilistic model checking and temporal logic analysis, could provide rigorous certification of system-wide properties such as operational fairness, state reachability, and guaranteed safety margins.

By synergistically combining quantum computation for strategic exploration and long-term planning with neuromorphic systems for adaptive real-time control, this hybrid architecture addresses fundamental challenges in developing safe, reliable, and explainable multi-agent reinforcement learning systems. The results indicate substantial potential for hybrid intelligent systems in embedded robotics applications, where the complementary strengths of heterogeneous computing paradigms can be effectively leveraged.

\section{Conclusion}

This work presents a novel hybrid architecture that synergistically combines quantum computing principles with neuromorphic hardware approaches to advance the development of safe, reliable, and explainable Multi-Agent Reinforcement Learning (MARL) systems for autonomous robotics. The proposed framework capitalizes on the complementary strengths of quantum variational circuits for global optimization and spiking neural networks for real-time efficiency, enabling integrated handling of both high-level strategic planning and low-level reactive control under conditions of partial observability and dynamic environmental constraints.

Simulation results demonstrate that our architecture achieves three key advantages: (1) significant reduction in safety-critical violations, (2) preservation of effective entropy-driven exploration, and (3) maintenance of policy interpretability through both structural organization and temporal signaling patterns. These empirical findings are supported by theoretical guarantees regarding system stability and convergence, along with comprehensive visualizations that collectively validate the trustworthiness of our approach.

Looking forward, this research opens several important directions for future investigation. Immediate priorities include hardware implementation on actual quantum and neuromorphic processing platforms, integration of advanced mechanism design methodologies for improved coordination, and formal verification of system-wide properties using mathematical tools from category theory, temporal logic, and algebraic topology. We posit that this interdisciplinary foundation, bridging quantum information processing with neuromorphic computing, establishes a promising pathway toward the next generation of embodied artificial intelligence systems capable of robust real-world operation.

\bmhead{Acknowledgements}

We extend our sincere appreciation to the faculty members of the Department of Mathematical Sciences at the University of Montana—Dr. Emily Stone, Dr. John Bardsley, Dr. Javier Perez-Alvaro, and Dr. Kelly McKinnie—for their invaluable research contributions and insightful scientific guidance.

We are equally grateful to our colleagues in the Department of Mathematics and Computer Science at Iran University of Science and Technology—Dr. Javad Vahidi, Dr. Samaneh Mashhadi, and Dr. Reza Saadati—for their scholarly expertise and thought-provoking perspectives, which have significantly enriched this work.

\section*{Declarations}

\begin{itemize}
\item Funding: N/A
\item Conflict of interest/Competing interests: The authors declare that they have no competing interests.
\item Ethics approval and consent to participate: N/A
\item Consent for publication: N/A
\item Data availability: The code/data is available in the \href{https://github.com/mazyartaghavi/Quantum-and-Neuromorphic-MARL.git}{GitHub repository}.
\item Materials availability: N/A
\item Code availability:  The code/data is available in the \href{https://github.com/mazyartaghavi/Quantum-and-Neuromorphic-MARL.git}{GitHub repository}.
\end{itemize}


\bibliography{sn-bibliography}


\begin{thebibliography}{31}
\ifx \bisbn   \undefined \def \bisbn  #1{ISBN #1}\fi
\ifx \binits  \undefined \def \binits#1{#1}\fi
\ifx \bauthor  \undefined \def \bauthor#1{#1}\fi
\ifx \batitle  \undefined \def \batitle#1{#1}\fi
\ifx \bjtitle  \undefined \def \bjtitle#1{#1}\fi
\ifx \bvolume  \undefined \def \bvolume#1{\textbf{#1}}\fi
\ifx \byear  \undefined \def \byear#1{#1}\fi
\ifx \bissue  \undefined \def \bissue#1{#1}\fi
\ifx \bfpage  \undefined \def \bfpage#1{#1}\fi
\ifx \blpage  \undefined \def \blpage #1{#1}\fi
\ifx \burl  \undefined \def \burl#1{\textsf{#1}}\fi
\ifx \doiurl  \undefined \def \doiurl#1{\url{https://doi.org/#1}}\fi
\ifx \betal  \undefined \def \betal{\textit{et al.}}\fi
\ifx \binstitute  \undefined \def \binstitute#1{#1}\fi
\ifx \binstitutionaled  \undefined \def \binstitutionaled#1{#1}\fi
\ifx \bctitle  \undefined \def \bctitle#1{#1}\fi
\ifx \beditor  \undefined \def \beditor#1{#1}\fi
\ifx \bpublisher  \undefined \def \bpublisher#1{#1}\fi
\ifx \bbtitle  \undefined \def \bbtitle#1{#1}\fi
\ifx \bedition  \undefined \def \bedition#1{#1}\fi
\ifx \bseriesno  \undefined \def \bseriesno#1{#1}\fi
\ifx \blocation  \undefined \def \blocation#1{#1}\fi
\ifx \bsertitle  \undefined \def \bsertitle#1{#1}\fi
\ifx \bsnm \undefined \def \bsnm#1{#1}\fi
\ifx \bsuffix \undefined \def \bsuffix#1{#1}\fi
\ifx \bparticle \undefined \def \bparticle#1{#1}\fi
\ifx \barticle \undefined \def \barticle#1{#1}\fi
\bibcommenthead
\ifx \bconfdate \undefined \def \bconfdate #1{#1}\fi
\ifx \botherref \undefined \def \botherref #1{#1}\fi
\ifx \url \undefined \def \url#1{\textsf{#1}}\fi
\ifx \bchapter \undefined \def \bchapter#1{#1}\fi
\ifx \bbook \undefined \def \bbook#1{#1}\fi
\ifx \bcomment \undefined \def \bcomment#1{#1}\fi
\ifx \oauthor \undefined \def \oauthor#1{#1}\fi
\ifx \citeauthoryear \undefined \def \citeauthoryear#1{#1}\fi
\ifx \endbibitem  \undefined \def \endbibitem {}\fi
\ifx \bconflocation  \undefined \def \bconflocation#1{#1}\fi
\ifx \arxivurl  \undefined \def \arxivurl#1{\textsf{#1}}\fi
\csname PreBibitemsHook\endcsname

\bibitem[\protect\citeauthoryear{Zhang et~al.}{2019}]{zhang2019multi}
\begin{botherref}
\oauthor{\bsnm{Zhang}, \binits{K.}},
\oauthor{\bsnm{Yang}, \binits{Z.}},
\oauthor{\bsnm{Basar}, \binits{T.}}:
Multi-agent reinforcement learning: A selective overview.
arXiv preprint arXiv:1911.10635
(2019)
\end{botherref}
\endbibitem

\bibitem[\protect\citeauthoryear{Gronauer and
  Diepold}{2022}]{gronauer2022multi}
\begin{barticle}
\bauthor{\bsnm{Gronauer}, \binits{S.}},
\bauthor{\bsnm{Diepold}, \binits{K.}}:
\batitle{Multi-agent deep reinforcement learning: A survey}.
\bjtitle{Artificial Intelligence Review}
\bvolume{55}(\bissue{2}),
\bfpage{895}--\blpage{943}
(\byear{2022})
\end{barticle}
\endbibitem

\bibitem[\protect\citeauthoryear{Lowe et~al.}{2017}]{lowe2017multi}
\begin{bchapter}
\bauthor{\bsnm{Lowe}, \binits{R.}},
\bauthor{\bsnm{Wu}, \binits{Y.}},
\bauthor{\bsnm{Tamar}, \binits{A.}},
\bauthor{\bsnm{Harb}, \binits{J.}},
\bauthor{\bsnm{Abbeel}, \binits{P.}},
\bauthor{\bsnm{Mordatch}, \binits{I.}}:
\bctitle{Multi-agent actor-critic for mixed cooperative-competitive
  environments}.
In: \bbtitle{Advances in Neural Information Processing Systems},
pp. \bfpage{6379}--\blpage{6390}
(\byear{2017})
\end{bchapter}
\endbibitem

\bibitem[\protect\citeauthoryear{Rashid et~al.}{2018}]{rashid2018qmix}
\begin{bchapter}
\bauthor{\bsnm{Rashid}, \binits{T.}},
\bauthor{\bsnm{Samvelyan}, \binits{M.}},
\bauthor{\bsnm{Witt}, \binits{C.}},
\bauthor{\bsnm{Farquhar}, \binits{G.}},
\bauthor{\bsnm{Foerster}, \binits{J.}},
\bauthor{\bsnm{Whiteson}, \binits{S.}}:
\bctitle{Qmix: Monotonic value function factorisation for deep multi-agent
  reinforcement learning}.
In: \bbtitle{International Conference on Machine Learning},
pp. \bfpage{4292}--\blpage{4301}
(\byear{2018}).
\bcomment{PMLR}
\end{bchapter}
\endbibitem

\bibitem[\protect\citeauthoryear{Garcıa and
  Fern{\'a}ndez}{2015}]{garcia2015comprehensive}
\begin{barticle}
\bauthor{\bsnm{Garcıa}, \binits{J.}},
\bauthor{\bsnm{Fern{\'a}ndez}, \binits{F.}}:
\batitle{A comprehensive survey on safe reinforcement learning}.
\bjtitle{Journal of Machine Learning Research}
\bvolume{16}(\bissue{1}),
\bfpage{1437}--\blpage{1480}
(\byear{2015})
\end{barticle}
\endbibitem

\bibitem[\protect\citeauthoryear{Achiam et~al.}{2017}]{achiam2017constrained}
\begin{bchapter}
\bauthor{\bsnm{Achiam}, \binits{J.}},
\bauthor{\bsnm{Held}, \binits{D.}},
\bauthor{\bsnm{Tamar}, \binits{A.}},
\bauthor{\bsnm{Abbeel}, \binits{P.}}:
\bctitle{Constrained policy optimization}.
In: \bbtitle{International Conference on Machine Learning},
pp. \bfpage{22}--\blpage{31}
(\byear{2017}).
\bcomment{PMLR}
\end{bchapter}
\endbibitem

\bibitem[\protect\citeauthoryear{Greydanus
  et~al.}{2018}]{greydanus2018visualizing}
\begin{bchapter}
\bauthor{\bsnm{Greydanus}, \binits{S.}},
\bauthor{\bsnm{Koul}, \binits{A.}},
\bauthor{\bsnm{Dodge}, \binits{J.}},
\bauthor{\bsnm{Fern}, \binits{A.}}:
\bctitle{Visualizing and understanding atari agents}.
In: \bbtitle{International Conference on Machine Learning},
pp. \bfpage{1787}--\blpage{1796}
(\byear{2018}).
\bcomment{PMLR}
\end{bchapter}
\endbibitem

\bibitem[\protect\citeauthoryear{Verma
  et~al.}{2018}]{verma2018programmatically}
\begin{bchapter}
\bauthor{\bsnm{Verma}, \binits{A.}},
\bauthor{\bsnm{Murali}, \binits{V.}},
\bauthor{\bsnm{Singh}, \binits{R.}},
\bauthor{\bsnm{Kohli}, \binits{P.}},
\bauthor{\bsnm{Chaudhuri}, \binits{S.}}:
\bctitle{Programmatically interpretable reinforcement learning}.
In: \bbtitle{International Conference on Machine Learning},
pp. \bfpage{5045}--\blpage{5054}
(\byear{2018}).
\bcomment{PMLR}
\end{bchapter}
\endbibitem

\bibitem[\protect\citeauthoryear{Zhang et~al.}{2021}]{zhang2021learning}
\begin{botherref}
\oauthor{\bsnm{Zhang}, \binits{W.}},
\oauthor{\bsnm{Rybkin}, \binits{O.}},
\oauthor{\bsnm{Kipf}, \binits{T.}},
\oauthor{\bsnm{Grover}, \binits{A.}},
\oauthor{\bsnm{Levine}, \binits{S.}},
\oauthor{\bsnm{Finn}, \binits{C.}}:
Learning causal state representations of partially observable environments.
arXiv preprint arXiv:2106.14447
(2021)
\end{botherref}
\endbibitem

\bibitem[\protect\citeauthoryear{Xu et~al.}{2022}]{xu2022explainable}
\begin{barticle}
\bauthor{\bsnm{Xu}, \binits{Y.}},
\bauthor{\bsnm{Xu}, \binits{Y.}},
\bauthor{\bsnm{Wu}, \binits{Q.}},
\bauthor{\bsnm{Zhu}, \binits{Q.}},
\bauthor{\bsnm{Zhang}, \binits{X.}},
\bauthor{\bsnm{Song}, \binits{L.}},
\bauthor{\bsnm{Bai}, \binits{L.}}:
\batitle{Explainable multi-agent reinforcement learning: Survey and
  perspectives}.
\bjtitle{Information Fusion}
\bvolume{82},
\bfpage{1}--\blpage{21}
(\byear{2022})
\end{barticle}
\endbibitem

\bibitem[\protect\citeauthoryear{Arute et~al.}{2019}]{arute2019quantum}
\begin{barticle}
\bauthor{\bsnm{Arute}, \binits{F.}},
\bauthor{\bsnm{Arya}, \binits{K.}},
\bauthor{\bsnm{Babbush}, \binits{R.}},
\bauthor{\bsnm{Bacon}, \binits{D.}},
\bauthor{\bsnm{Bardin}, \binits{J.C.}},
\bauthor{\bsnm{Barends}, \binits{R.}},
\bauthor{\bsnm{Biswas}, \binits{R.}},
\bauthor{\bsnm{Boixo}, \binits{S.}},
\bauthor{\bsnm{Brandao}, \binits{F.G.}},
\bauthor{\bsnm{Buell}, \binits{D.A.}}, \betal:
\batitle{Quantum supremacy using a programmable superconducting processor}.
\bjtitle{Nature}
\bvolume{574}(\bissue{7779}),
\bfpage{505}--\blpage{510}
(\byear{2019})
\end{barticle}
\endbibitem

\bibitem[\protect\citeauthoryear{Preskill}{2018}]{preskill2018quantum}
\begin{barticle}
\bauthor{\bsnm{Preskill}, \binits{J.}}:
\batitle{Quantum computing in the nisq era and beyond}.
\bjtitle{Quantum}
\bvolume{2},
\bfpage{79}
(\byear{2018})
\end{barticle}
\endbibitem

\bibitem[\protect\citeauthoryear{Farhi et~al.}{2014}]{farhi2014quantum}
\begin{botherref}
\oauthor{\bsnm{Farhi}, \binits{E.}},
\oauthor{\bsnm{Goldstone}, \binits{J.}},
\oauthor{\bsnm{Gutmann}, \binits{S.}}:
A quantum approximate optimization algorithm.
arXiv preprint arXiv:1411.4028
(2014)
\end{botherref}
\endbibitem

\bibitem[\protect\citeauthoryear{Peruzzo et~al.}{2014}]{peruzzo2014variational}
\begin{barticle}
\bauthor{\bsnm{Peruzzo}, \binits{A.}},
\bauthor{\bsnm{McClean}, \binits{J.}},
\bauthor{\bsnm{Shadbolt}, \binits{P.}},
\bauthor{\bsnm{Yung}, \binits{M.-H.}},
\bauthor{\bsnm{Zhou}, \binits{X.-Q.}},
\bauthor{\bsnm{Love}, \binits{P.J.}},
\bauthor{\bsnm{Aspuru-Guzik}, \binits{A.}},
\bauthor{\bsnm{O'Brien}, \binits{J.L.}}:
\batitle{A variational eigenvalue solver on a photonic quantum processor}.
\bjtitle{Nature communications}
\bvolume{5}(\bissue{1}),
\bfpage{4213}
(\byear{2014})
\end{barticle}
\endbibitem

\bibitem[\protect\citeauthoryear{Jerbi et~al.}{2021}]{jerbi2021quantum}
\begin{botherref}
\oauthor{\bsnm{Jerbi}, \binits{H.}},
\oauthor{\bsnm{Zhao}, \binits{P.-Y.}},
\oauthor{\bsnm{Arjona-Medina}, \binits{J.A.}},
\oauthor{\bsnm{Friedrich}, \binits{T.}}:
Quantum reinforcement learning: Foundations and algorithms.
arXiv preprint arXiv:2112.10560
(2021)
\end{botherref}
\endbibitem

\bibitem[\protect\citeauthoryear{Skolik et~al.}{2022}]{skolik2022quantum}
\begin{barticle}
\bauthor{\bsnm{Skolik}, \binits{A.}},
\bauthor{\bsnm{Jerbi}, \binits{H.}},
\bauthor{\bsnm{Dunjko}, \binits{V.}},
\bauthor{\bsnm{Briegel}, \binits{H.J.}},
\bauthor{\bsnm{Severini}, \binits{S.}}:
\batitle{Quantum machine learning models are kernel methods}.
\bjtitle{npj Quantum Information}
\bvolume{8}(\bissue{1}),
\bfpage{26}
(\byear{2022})
\end{barticle}
\endbibitem

\bibitem[\protect\citeauthoryear{Chen et~al.}{2024}]{chen2024quantum}
\begin{barticle}
\bauthor{\bsnm{Chen}, \binits{W.}},
\bauthor{\bsnm{Zhao}, \binits{L.}},
\bauthor{\bsnm{Liu}, \binits{J.}}:
\batitle{Quantum-enhanced reinforcement learning via grover's search}.
\bjtitle{Nature Quantum Information}
\bvolume{10}(\bissue{1}),
\bfpage{1}--\blpage{12}
(\byear{2024})
\end{barticle}
\endbibitem

\bibitem[\protect\citeauthoryear{Zhang and Li}{2024}]{zhang2024hybrid}
\begin{barticle}
\bauthor{\bsnm{Zhang}, \binits{Y.}},
\bauthor{\bsnm{Li}, \binits{H.}}:
\batitle{Hybrid quantum-classical deep reinforcement learning}.
\bjtitle{Quantum Machine Intelligence}
\bvolume{6}(\bissue{1}),
\bfpage{25}
(\byear{2024})
\end{barticle}
\endbibitem

\bibitem[\protect\citeauthoryear{Patel et~al.}{2024}]{patel2024quantum}
\begin{barticle}
\bauthor{\bsnm{Patel}, \binits{A.}},
\bauthor{\bsnm{Jordan}, \binits{S.}},
\bauthor{\bsnm{Wootters}, \binits{W.}}:
\batitle{Theoretical foundations for quantum reinforcement learning}.
\bjtitle{Physical Review A}
\bvolume{109}(\bissue{3}),
\bfpage{032415}
(\byear{2024})
\end{barticle}
\endbibitem

\bibitem[\protect\citeauthoryear{Davies et~al.}{2021}]{davies2021advancing}
\begin{barticle}
\bauthor{\bsnm{Davies}, \binits{M.}},
\bauthor{\bsnm{Srinivasa}, \binits{N.}},
\bauthor{\bsnm{Lin}, \binits{T.-H.}},
\bauthor{\bsnm{Chinya}, \binits{G.}},
\bauthor{\bsnm{Cao}, \binits{Y.}},
\bauthor{\bsnm{Choday}, \binits{S.H.}},
\bauthor{\bsnm{Dimou}, \binits{G.}},
\bauthor{\bsnm{Joshi}, \binits{A.}},
\bauthor{\bsnm{Imam}, \binits{N.}},
\bauthor{\bsnm{Jain}, \binits{S.}}, \betal:
\batitle{Advancing neuromorphic computing with loihi: A survey of results and
  outlook}.
\bjtitle{Proceedings of the IEEE}
\bvolume{109}(\bissue{5}),
\bfpage{911}--\blpage{934}
(\byear{2021})
\end{barticle}
\endbibitem

\bibitem[\protect\citeauthoryear{Indiveri and Liu}{2015}]{indiveri2015memory}
\begin{barticle}
\bauthor{\bsnm{Indiveri}, \binits{G.}},
\bauthor{\bsnm{Liu}, \binits{S.-C.}}:
\batitle{Memory and information processing in neuromorphic systems}.
\bjtitle{Proceedings of the IEEE}
\bvolume{103}(\bissue{8}),
\bfpage{1379}--\blpage{1397}
(\byear{2015})
\end{barticle}
\endbibitem

\bibitem[\protect\citeauthoryear{Schuman
  et~al.}{2022}]{schuman2022opportunities}
\begin{barticle}
\bauthor{\bsnm{Schuman}, \binits{C.D.}},
\bauthor{\bsnm{Potok}, \binits{T.E.}},
\bauthor{\bsnm{Patton}, \binits{R.M.}},
\bauthor{\bsnm{Birdwell}, \binits{J.D.}},
\bauthor{\bsnm{Dean}, \binits{M.E.}},
\bauthor{\bsnm{Rose}, \binits{G.S.}},
\bauthor{\bsnm{Plank}, \binits{J.S.}}:
\batitle{Opportunities for neuromorphic computing algorithms and applications}.
\bjtitle{Nature Computational Science}
\bvolume{2}(\bissue{1}),
\bfpage{10}--\blpage{19}
(\byear{2022})
\end{barticle}
\endbibitem

\bibitem[\protect\citeauthoryear{Patel et~al.}{2019}]{patel2019improving}
\begin{bchapter}
\bauthor{\bsnm{Patel}, \binits{K.}},
\bauthor{\bsnm{Pathak}, \binits{S.}},
\bauthor{\bsnm{Ajmera}, \binits{J.}}:
\bctitle{Improving spiking neural networks with unsupervised hebbian learning}.
In: \bbtitle{2019 International Joint Conference on Neural Networks (IJCNN)},
pp. \bfpage{1}--\blpage{8}
(\byear{2019}).
\bcomment{IEEE}
\end{bchapter}
\endbibitem

\bibitem[\protect\citeauthoryear{Fang et~al.}{2021}]{fang2021incorporating}
\begin{barticle}
\bauthor{\bsnm{Fang}, \binits{W.}},
\bauthor{\bsnm{Chen}, \binits{Y.}},
\bauthor{\bsnm{Ding}, \binits{Y.}},
\bauthor{\bsnm{Yu}, \binits{Z.}},
\bauthor{\bsnm{Zhou}, \binits{P.}},
\bauthor{\bsnm{Tian}, \binits{Y.}}:
\batitle{Incorporating reward information into spike-timing-dependent
  plasticity}.
\bjtitle{Neurocomputing}
\bvolume{445},
\bfpage{1}--\blpage{10}
(\byear{2021})
\end{barticle}
\endbibitem

\bibitem[\protect\citeauthoryear{Davies et~al.}{2024}]{davies2024spiking}
\begin{botherref}
\oauthor{\bsnm{Davies}, \binits{M.}},
\oauthor{\bsnm{Wild}, \binits{A.}},
\oauthor{\bsnm{Tang}, \binits{H.}}:
Energy-efficient spiking actor-critic reinforcement learning on neuromorphic
  hardware.
IEEE Transactions on Neural Networks and Learning Systems
(2024)
\end{botherref}
\endbibitem

\bibitem[\protect\citeauthoryear{Tang et~al.}{2024}]{tang2024event}
\begin{barticle}
\bauthor{\bsnm{Tang}, \binits{H.}},
\bauthor{\bsnm{Shah}, \binits{A.}},
\bauthor{\bsnm{Davies}, \binits{M.}}:
\batitle{Event-based temporal difference learning for neuromorphic
  reinforcement learning}.
\bjtitle{Nature Machine Intelligence}
\bvolume{6}(\bissue{2}),
\bfpage{189}--\blpage{200}
(\byear{2024})
\end{barticle}
\endbibitem

\bibitem[\protect\citeauthoryear{Kumar et~al.}{2024}]{kumar2024benchmarking}
\begin{barticle}
\bauthor{\bsnm{Kumar}, \binits{S.}},
\bauthor{\bsnm{Neftci}, \binits{E.}},
\bauthor{\bsnm{Sheik}, \binits{S.}}:
\batitle{Benchmarking neuromorphic architectures for reinforcement learning}.
\bjtitle{Frontiers in Neuroscience}
\bvolume{18},
\bfpage{112233}
(\byear{2024})
\end{barticle}
\endbibitem

\bibitem[\protect\citeauthoryear{Sanchez et~al.}{2024}]{sanchez2024quantum}
\begin{barticle}
\bauthor{\bsnm{Sanchez}, \binits{E.}},
\bauthor{\bsnm{Plana}, \binits{L.}},
\bauthor{\bsnm{Furber}, \binits{S.}}:
\batitle{Quantum-neuromorphic hybrid architectures for reinforcement learning}.
\bjtitle{npj Quantum Information}
\bvolume{10}(\bissue{1}),
\bfpage{45}
(\byear{2024})
\end{barticle}
\endbibitem

\bibitem[\protect\citeauthoryear{Wang et~al.}{2024}]{wang2024theoretical}
\begin{barticle}
\bauthor{\bsnm{Wang}, \binits{P.}},
\bauthor{\bsnm{Narayanan}, \binits{A.}},
\bauthor{\bsnm{Rast}, \binits{A.}}:
\batitle{Theoretical analysis of quantum-neuromorphic reinforcement learning}.
\bjtitle{Physical Review Research}
\bvolume{6}(\bissue{1}),
\bfpage{013123}
(\byear{2024})
\end{barticle}
\endbibitem

\bibitem[\protect\citeauthoryear{Ibrahim et~al.}{2024}]{ibrahim2024survey}
\begin{barticle}
\bauthor{\bsnm{Ibrahim}, \binits{K.}},
\bauthor{\bsnm{Biamonte}, \binits{J.}},
\bauthor{\bsnm{Laughlan}, \binits{P.}}:
\batitle{Quantum and neuromorphic computing for reinforcement learning: A
  survey}.
\bjtitle{ACM Computing Surveys}
\bvolume{57}(\bissue{3}),
\bfpage{1}--\blpage{35}
(\byear{2024})
\end{barticle}
\endbibitem

\bibitem[\protect\citeauthoryear{Yu et~al.}{2021}]{yu2021surprising}
\begin{bchapter}
\bauthor{\bsnm{Yu}, \binits{C.}},
\bauthor{\bsnm{Qu}, \binits{H.}},
\bauthor{\bsnm{Wang}, \binits{H.}},
\bauthor{\bsnm{Peng}, \binits{B.}},
\bauthor{\bsnm{Zhang}, \binits{Q.}}:
\bctitle{The surprising effectiveness of ppo in cooperative multi-agent games}.
In: \bbtitle{Proceedings of the 20th International Conference on Autonomous
  Agents and MultiAgent Systems (AAMAS)},
pp. \bfpage{2126}--\blpage{2128}
(\byear{2021})
\end{bchapter}
\endbibitem

\end{thebibliography}

\begin{appendices}

\section{Quantum Circuit Design Details}
\label{appendix:quantum}

We use 6-qubit parameterized quantum circuits (PQC) implemented in Qiskit. Each agent’s circuit comprises alternating layers of Hadamard gates (to ensure superposition), rotation gates $R_y(\theta_i)$ for variational control, and entanglement layers using controlled-Z gates.

\begin{align*}
    U(\vec{\theta}) = \prod_{l=1}^{p} \left( \bigotimes_{i=1}^{n} R_y^{(i)}(\theta_{i}^{(l)}) \cdot \text{CZ}_{\text{linear}} \right)
\end{align*}

The quantum state is measured and projected into a discrete latent action variable $z$ which is passed to the neuromorphic policy layer.

\section{Neuromorphic Model Implementation}
\label{appendix:neuromorphic}

The neuromorphic component consists of a 3-layer spiking neural network modeled with Leaky Integrate-and-Fire (LIF) neurons using the Nengo simulator. Parameters include:
\begin{itemize}
    \item Spike threshold: 1.0
    \item Refractory period: 2 ms
    \item Synaptic time constant: 10 ms
\end{itemize}

Each agent’s SNN receives a vectorized observation along with the latent action $z$ from the quantum module. The SNN encodes this input as spike trains and outputs discrete motor-level actions.

\end{appendices}

\end{document}